%% file: template.tex
\documentclass{WileyMSP-template}
\usepackage{multirow}
\usepackage{lineno}
\usepackage{amsmath}
\usepackage{xcolor}

\newcommand{\red}[1]{{\color{black}#1}}

\begin{document}

\pagestyle{fancy}
\rhead{\includegraphics[width=2.5cm]{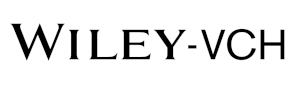}}

\title{Room-temperature exceptional-point-driven polariton lasing from perovskite metasurface}

\maketitle


\author{M.A.~Masharin}
\author{A.K.~Samusev}
\author{A.A.~Bogdanov}
\author{I.V.~Iorsh}
\author{H.V.~Demir}
\author{S.V.~Makarov*}


\dedication{}

\begin{affiliations}
M.A.~Masharin, H.V.~Demir\\
Institute of Materials Science and Nanotechnology and National Nanotechnology Research Center and Electronics Engineering and National Nanotechnology Research Center, Department of Electrical, Department of Physics, Bilkent University, Ankara, 06800, Turkey\\

M.A.~Masharin, A.K.~Samusev, A.A.~Bogdanov, I.V.~Iorsh, S.V.~Makarov\\
ITMO University, School of Physics and Engineering, St. Petersburg, 197101, Russia\\
Email Address: s.makarov@metalab.ifmo.ru\\

A.K.~Samusev\\
Experimentelle Physik 2, Technische Universit\"at Dortmund, 44227 Dortmund, Germany\\

A.A.~Bogdanov, S.V.~Makarov\\
Qingdao Innovation and Development Center, Harbin Engineering University, Qingdao 266000, Shandong, China\\

I.V.~Iorsh\\
Department of Physics, Bar-Ilan University, Ramat Gan, 52900, Israel\\

H.V.~Demir\\
LUMINOUS! Center of Excellence for Semiconductor Lighting and Displays, School of Electrical and Electronic Engineering, School of Physical and Materials Sciences, School of Materials Science and Engineering, Nanyang Technological University, 639798, Singapore\\

\end{affiliations}



\begin{abstract}

Excitons in lead bromide perovskites exhibit high binding energy and high oscillator strength, allowing for a strong light-matter coupling regime in the perovskite-based cavities localizing photons at the nanoscale. 
This opens up the way for the realization of exciton-polariton Bose-Einstein condensation and polariton lasing at room temperature -- the inversion-free low-threshold stimulated emission. 
However, polariton lasing in perovskite planar photon cavities without Bragg mirrors has not yet been observed and proved experimentally. 
In this work, we employ perovskite metasurface, fabricated with nanoimprint lithography, supporting so-called exceptional points to demonstrate the room-temperature polariton lasing. The exceptional points in exciton-polariton dispersion of the metasurface appear upon optically pumping in the nonlinear regime in the spectral vicinity of a symmetry-protected bound state in the continuum providing high mode confinement with the enhanced local density of states beneficial for polariton condensation.
The observed lasing emission possesses high directivity  with a divergence angle of around 1$^\circ$ over one axis. The employed nanoimprinting approach for solution-processable large-scale polariton lasers is compatible with various planar photonic platforms suitable for on-chip integration.

\end{abstract}


\section{Introduction}

Exciton-polaritons are bosonic hybrid part-light, part-matter quasiparticles, which have attracted tremendous attention \red{thanks} to a set of remarkable properties.~\cite{sanvitto2012exciton} The photon fraction provides a small effective mass and long decoherence time, whereas the excitonic fraction enables strong optical nonlinearity caused by the Coulomb interaction between quasiparticles. The unique combination of the properties allows the realization of Bose-Einstein condensation -- the state, where bosons \red{are accumulated} in one quantum state exhibiting collective coherence on a macroscopic scale.\cite{deng2002condensation,kasprzak2006bose} In the case of non-equilibrium polariton systems, it turns to the low-threshold polariton lasing, which does not require inversion of population. \cite{deng2003polariton} The most widely used photon cavity for polariton systems is distributed Bragg reflectors since such a cavity supports high-quality optical states with parabolic dispersion, which can be measured directly with the angle-resolved spectroscopy method.\cite{weisbuch1992observation} \red{The} strong light-matter coupling regime and polariton lasing \red{were} already demonstrated for many materials,\cite{kasprzak2006bose,deng2010exciton,estrecho2019direct} even at room temperature.\cite{christopoulos2007room,su2017room,su2018room,betzold2019coherence} However, polariton lasers can be also realized in other planar cavities, such as metasurfaces, which have recently attracted high attention thanks to their unique properties.\cite{koshelev2019meta}

Metasurfaces can be classified by their functionality and design.\cite{overvig2022diffractive} If the operating wavelength is longer or comparable with the metasurface period, light can propagate along the metasurface, providing a non-local response.\cite{overvig2022diffractive} In such non-local metasurfaces, it is possible to realize high-Q resonances with large optical confinement without exploiting Bragg mirrors. The high-Q resonances originating from the destructive interference of counterpropagating waves are called bound states in the continuum (BICs). \cite{kravtsov2020nonlinear,koshelev2018strong,koshelev2019meta} By \red{varying} the metasurface design it is possible to precisely control the spectral position and radiative lifetime of the optical resonance. In \red{opposite to} Bragg mirrors, \red{a} planar metasurface can be fabricated with substantially simplified techniques, such as nanoimprint lithography. \red{Moreover,} if the metasurface design provides the BICs strongly coupled with the exciton resonance, \red{polariton condensation state in a nonlinear regime can be realized}.\cite{kravtsov2020nonlinear,ardizzone2022polariton}

However, there \red{is} another specific state in metasurfaces for \red{the} efficient polariton accumulation. When two hybrid optical modes are not split but degenerated at some point in the dispersion curve, it \red{might} result in the emergence of exceptional points (EPs).\cite{miri2019exceptional} These exotic states are very sensitive to parameter variations and were also observed in a rich variety of systems including whispering gallery mode resonators,\cite{peng2016chiral} metasurfaces,\cite{miri2019exceptional} and coupled waveguides.\cite{goldzak2018light} EPs can enhance lasing emission and allow to control \red{the} lasing directivity.\cite{doronin2019lasing,zyablovsky2021exceptional,peng2016chiral} Recently, it was also shown that EPs provide the enhanced local density of the states (LDOS), scaling with \red{a} square of the Purcell factor.\cite{lu2020engineering,pick2017general} If EPs appear on a polariton branch, this causes a strong enhancement of the boson scattering probability to this state, which can lead to the polariton accumulation in a nonlinear regime, resulting in the polariton lasing.\cite{zhang2022electric,shan2022brightening}

In turn, halide perovskites are one of the most promising candidates \red{to be} an active medium for \red{the} metasurface-based polariton lasers. This group of materials \red{was proved} to be a promising photo-active \red{material} for photovoltaics and optoelectronics \cite{jena2019halide,zhang2019photonics}. Moreover, its outstanding physical properties such as relatively high exciton binding energy and oscillator strength~\cite{su2021perovskite}, as well as high defect tolerance~\cite{huang2017lead}, allow for realizing a strong light-matter coupling regime in photonic cavities and even non-equilibrium polariton Bose-Einstein condensation at room temperature.\cite{su2020observation,su2018room,su2017room,shang2020role,peng2022room,tao2022halide} Moreover, solution-based methods of synthesis, combined with a high refractive index \red{of the synthesized materials}, and \red{their} scalable nanostructuring techniques, such as direct laser ablation,\cite{zhizhchenko2021direct} nanoimprint \red{lithogrpahy},\cite{makarov2017multifold} and self-assembly methods\cite{feng2021all}, open the \red{way} to apply them in planar metasurfaces, supporting high-Q polariton states \red{and} avoiding vertical Bragg cavities.\cite{feng2021all,al2022strong,as2022mie,dang2022realization} Although the perovskites were used in different laser \red{designs}\cite{veldhuis2016perovskite} including surface-emitting distributed feedback lasers,\cite{pourdavoud2018distributed,huang2020ultrafast,palatnik2021control} while polariton lasing in \red{halide} perovskite metasurfaces has not been yet observed experimentally.

\begin{figure}[t!]
\centering
\center{\includegraphics[width=0.8\linewidth]{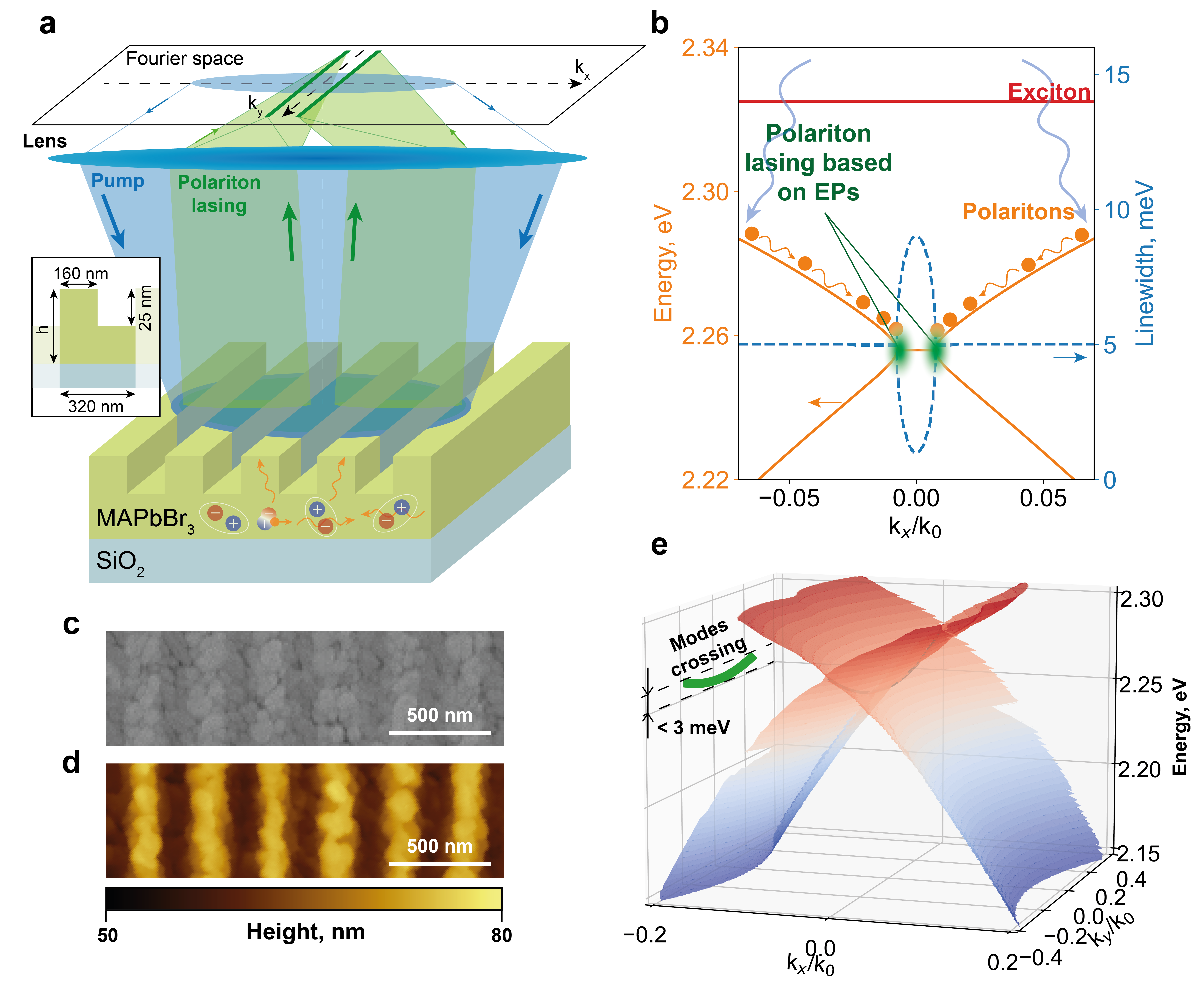}}
\caption{Concept of the polariton lasing enabled by the exceptional points in the perovskite metasurface. (a) Sketch of polariton lasing enhanced by EPs in perovskite metasurface\red{, where} femtosecond laser pumps the MAPbBr$_3$ \red{metasurface} \red{and} leaky modes strongly coupled to the exciton resonance, resulting in exciton-polaritons, \red{cause} the polariton lasing, \red{which is observable in Fourier space of the optical setup}. \red{The} stimulated emission arises at the in-plane wavevector $k_x$ close to zero in a wide range of $k_y$. (b) Sketch of the exciton-polariton dispersion \red{for} the leaky modes in \red{the} perovskite metasurface. The solid red line shows the exciton resonance, \red{and} the orange line shows the polariton dispersion. Blue dashed lines represent the linewidth of the polariton branches. The blue arrows represent carrier relaxation after the non-resonant pump into \red{the} polariton states. \red{The} range arrows illustrate the stimulated polariton relaxation into the localized state. \red{The} green ellipses correspond to the polariton lasing, based on exceptional points. (c) SEM and (d) AFM images of the fabricated perovskite metasurface. The geometrical parameters obtained from the AFM and SEM data are shown in the inset of (a).
(e) The dispersion surface of the studied metasurface calculated \red{by the} Fourier Modal Method. \red{Pseudo-colors correspond} to the energy. The black solid line shows the crossing of the counter-propagating modes at $k_x/k_0$ = 0 for various $k_y/k_0$. Its projection is shown with the green line and is limited to the angles $k_x/k_0$ $\approx$0.2, used in the experimental studies. The energy variation in the modes crossing does not exceed 3~meV in the considered range of $k_y/k_0$.}
\label{fig1}
\end{figure}

In this work, we experimentally demonstrate room-temperature exciton-polariton lasing in the perovskite non-local metasurface (see the design in Fig.~\ref{fig1}a). The polariton lasing is based on the EPs that appear in the spectral vicinity of a symmetry-protected BIC under a non-resonant optical pump in the nonlinear regime (Fig.~\ref{fig1}b). The observed lasing emission is characterized by unidirectional propagation along the metasurface periodicity direction and broad distribution along the perpendicular axis in the range of 11 degrees from the normal. Thanks to the dispersion of polariton modes in the perovskite metasurface, the EPs turn into curves in the Fourier plane, resulting in the observed lasing emission directivity.

\section{Results}
\subsection{Fabrication and characterization of the perovskite metasurface}

One of the most important advantages of perovskites is the solution methods of synthesis, which provide the low-cost fabrication of these materials and make them very promising for real-world applications.\cite{jeon2014solvent} Also, halide perovskites are known as an admirable material for nanoimprint lithography due to their crystal lattice softness, which is also very suitable for further scaling. \cite{makarov2017multifold,masharin2022polaron} We use both techniques for the fabrication of MAPbBr$_3$ metasurface, which demonstrate the high quality of the morphology and resulting optical modes (see Methods \red{and Supplementary Information (SI) }for details).

The perovskite metasurface represents a periodic grating, controlled by the mold, with a period of 320~nm, which is determined from the scanning electron microscopy (SEM) image, shown in Fig. \ref{fig1}c. The grating ridge height is around 20~nm, the comb width is around 160~nm, and the full sample thickness from the substrate to the comb top is \red{around} 75~nm, \red{according to} the atomic force microscopy (AFM) measurements shown in Fig.~\ref{fig1}d. The extracted grating profiles along the x-axis are shown in Fig. S2 in SI. We also fabricate the sample with a thickness of 65~nm by increasing the spin-coating rotation speed. It is observed that perovskite grains in the nanoimprinted structure are not deformed or damaged and the resulting structure possesses high crystallinity. According to our observations, due to the room-temperature nanoimprinting process applied during the intermediate phase, the crystallization process continues in the mold geometry during the nanoimprint process. Moreover, after the annealing, the structure stiffens and cannot be modified by the polycarbonate mold at room temperature.

Based on the extracted perovskite non-local metasurface geometry parameters and the MAPbBr$_3$ refractive index, studied before,\cite{alias2016optical} we estimate the wavevectors of leaky mode resonances with linear polarization co-direct with a groove of metasurface (TE polarization) as a function of the photon energy by the Fourier Modal Method (FMM)~\cite{li1997new} for the sample with a thickness of 75~nm. As the estimations well correlate with the experimentally measured angle-resolved reflection spectrum at $k_y/k_0$ = 0 (Fig. S3 in SI), we simulate the full dispersion surface of the perovskite metasurface shown in Fig. \ref{fig1}e (see section 1 of SI for details). In the current geometry, \red{the} modes with opposite x-components of the group velocities intersect at the curve, located in the plane of $k_x/k_0$ = 0. Note that, in the range of $|k_y/k_0|< 0.2$, spectral positions of the mode intersection points, shown as a green line in Fig. \ref{fig1}e, vary in the range of 3~meV. As the mode linewidth of $\approx$ 6~meV exceeds this value, the intersection set of points represents a straight line without any energy dependence. Also, the leaky modes are studied in the region of the exciton resonance and, therefore, are affected by it (Fig. S4 in SI). The curvature of the leaky mode at constant $k_y/k_0$ from the linear dependence to the exciton resonance asymptote is a sign of exciton-polariton behavior.\cite{hopfield1958theory}

\subsection{Strong light-matter coupling regime in perovskite metasurface}

\begin{figure}[t!]
\centering
\center{\includegraphics[width=0.8\linewidth]{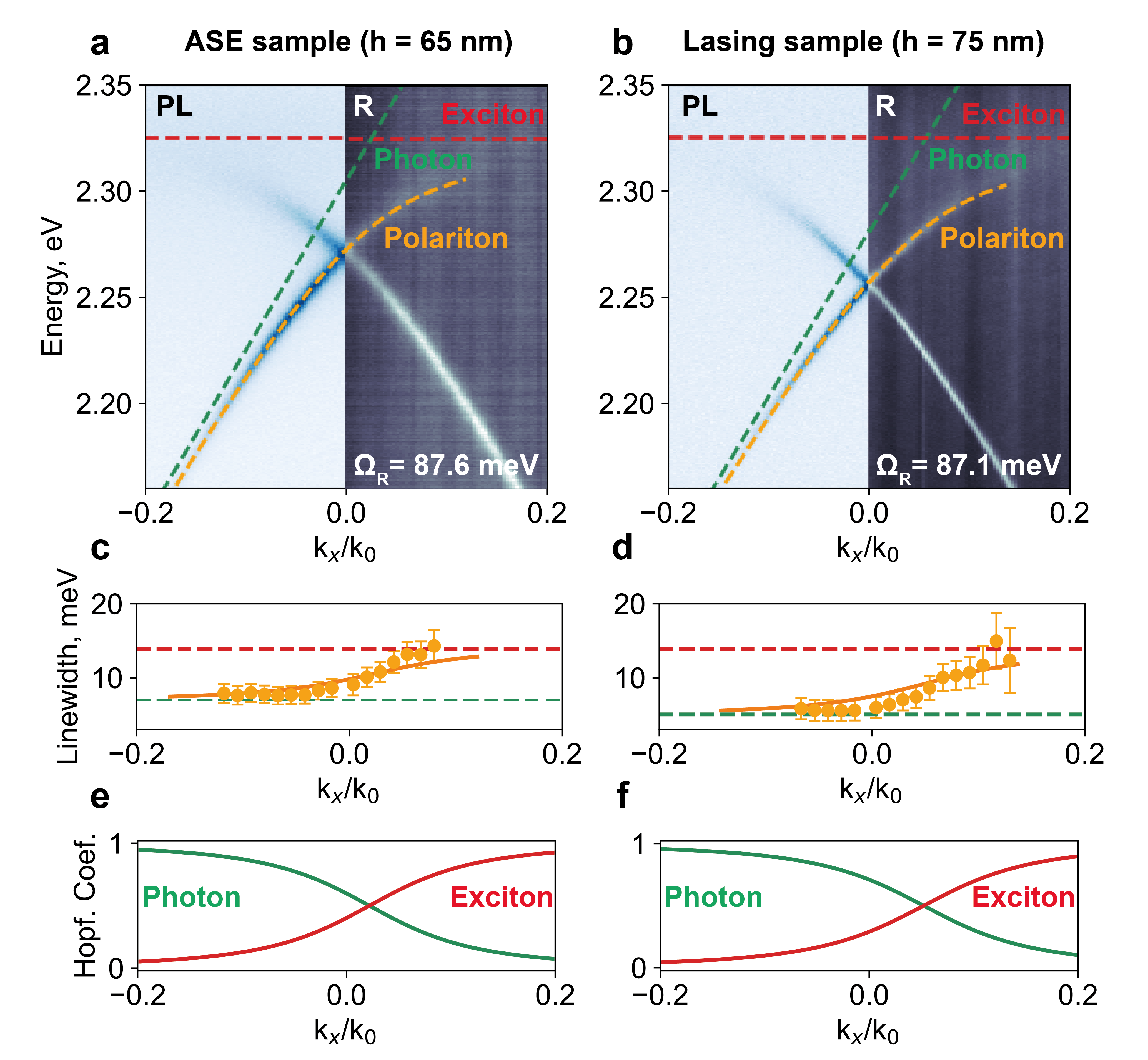}}
\caption{Fitting of the exciton-polariton branch for the samples without and with exceptional points, respectively. (a,b) Measured angle-resolved reflection and PL spectra for the ASE sample and lasing sample. 
Red dashed lines correspond to the exciton resonance, green dashed lines correspond to the estimated uncoupled photon cavity dispersion, and orange dashed lines correspond to the exciton-polariton dispersions, estimated by the two-coupled oscillators model.\cite{hopfield1958theory} The resulting Rabi splitting is equal to 87.6 and 87.1~meV, respectively (c,d) Linewidths extracted from experimental data are shown by orange dots. Estimated exciton linewidth is shown by red dashed lines and uncoupled cavity photon linewidths are shown by green dashed lines. The solid orange lines show the result of the two-coupled model fitting. \red{The error bars show the root-mean-squared error of the extracted mode linewidths} (e,f) Estimated Hopfield coefficients based on the fitted two-coupled model parameters. The red line corresponds to the exciton fraction $|X|^2$ and the green line corresponds to the photon fraction $|C|^2$ (See Section 2 in SI for details).}
\label{fig2}
\end{figure}

In order to confirm the strong light-matter coupling regime, we perform angle-resolved reflectance and photoluminescence measurements (see Methods for details). The angle-resolved measurements at \textit{$k_y/k_0$} = 0 in TE polarization are performed for two samples with thicknesses of 65 and 75~nm, but having identical pitch and modulation depth, shown in Figs.~\ref{fig2}a and 2b. The samples are labeled as an "ASE sample" and "lasing sample", respectively, as they provide spectrally broad and narrow stimulated emission, respectively, under \red{a} non-resonant femtosecond pump. The difference in the sample thickness causes the variation in the real $E_{LP}(k_x)$ and imaginary $\gamma_{LP}(k_x)$ parts of leaky modes dispersions (Figs. \ref{fig2}a and \ref{fig2}b) and linewidths (Figs. \ref{fig2}c and \ref{fig2}d). Both dispersions demonstrate the curvature of the mode near the exciton resonance, which is the sign of the exciton-polaritons, as was mentioned above. In the absence of the excitonic resonance, the leaky mode has nearly linear behavior as a function of $k_x/k_0$ in the spectral range of interest and can be considered as an uncoupled cavity photon (see Fig. S4 in SI for details). The observed polariton dispersion can be described by the two-coupled oscillator model\cite{hopfield1958theory}:
\begin{equation}
    E_{LP}(k_x) = \frac{{E}_{X} + {E}_{C}(k_x)}{2} - i\frac{{\gamma}_{X} + {\gamma}_{C}}{2} - \frac{1}{2}\sqrt{\left[({E}_{C}(k_x) - {E}_{X}) - i(\gamma_C - \gamma_X)\right]^2 + 4g_0^2}, 
    \label{CoupledOscillator}
\end{equation}
there $E_{LP}(k_x)$ stands for the complex energy dispersion of the lower polariton branch; ${E}_{X}$ and $\gamma_X$ stand for the real and imaginary parts of the exciton resonance, respectively; ${E}_{C}(k_x)$ and $\gamma_{C}$ stand for the dispersion and linewidth of the uncoupled cavity photon mode, respectively; and $g_0$ stands for the light-matter coupling coefficient. It should be noted, that we do not observe the upper polariton branch due to the strong optical absorption in the spectral region higher than the exciton resonance.\cite{sestu2015absorption} Nevertheless, Rabi splitting, which is identified as a difference between lower and upper polariton branches in the intersection point $E_X = E_C(k_x)$ can be estimated as

\begin{equation}
    \Omega_R = \sqrt{4g_0^2 - (\gamma_C - \gamma_X)^2},
    \label{OmegaR}
\end{equation}

If the coupling coefficient exceeds the half-difference of the linewidths (i.e., $g_0 > (\gamma_C - \gamma_X)/2$) and the value of Rabi splitting exceeds the half-sum (i.e., $\Omega_R > (\gamma_C + \gamma_X)/2$) strong light-matter coupling regime appears.\cite{hopfield1958theory}

In order to estimate the coupling coefficient $g_0$ and check if the system is in a strong light-matter coupling regime, we extract the leaky mode dispersion and linewidths $E_{LP}(k_x)$ from the measured angle-resolved PL spectra by the fitting of the mode resonances at each $k_x/k_0$ by the Lorentz peak function. As uncoupled photon cavity mode dispersion $E_C(k_x)$ in the metasurface is considered to have linear behavior with respect to the $k_x/k_0$, we estimate it by the linear approximation of the leaky mode in the spectral region far from the exciton resonance (1.9-2.0~eV). The linewidth $\gamma_C$ is estimated from the leaky mode linewidth in the same spectral region and is equal to 7~meV for the ASE sample and 5~meV for the lasing sample. The coupling coefficient $g_0$ is determined as an optimized parameter in the fitting by the two-coupled oscillator model. In the same way, the exciton resonance $E_X$ and linewidth $\gamma_x$ are also determined as optimized parameters and well correspond to the previous estimations.\cite{soufiani2015polaronic,shi2020exciton} It should be noted that this approach of the exciton level estimation well predicts the exciton resonance in comparison with other methods.\cite{masharin2022room} The result of the complex mode dispersion fitting by the two-coupled oscillator model is shown in Figs. \ref{fig2}a-d. The real part of the $E_{LP}(k_x)$, shown by the dashed lines in Figs. \ref{fig2}a-b well corresponds to the experimental data, as well as the imaginary part shown in Figs. \ref{fig2}c-d. The coupling coefficient $g_0$ is estimated to be equal to 43.8~meV for both samples, but $\Omega_R$ is 87.6~meV for the ASE sample and 87.1~meV for the lasing sample. As the coupling coefficient depends on the material properties and electric field localization, it is the same for both samples, but the difference is in the photon cavity linewidth, which causes the difference in $\Omega_R$. As for both samples, $g_0$ exceeds the exciton and photon linewidths half-difference, which is around 7 and 9~meV, as well as $\Omega_R$ is larger than the half-sum, which is around 9.5 and 10.5~meV, we confirm that in the studied perovskite metasurface, \red{the} strong light-matter coupling regime is \red{achieved}.

As exciton-polaritons are hybrid light-matter quasiparticles, there can be estimated a fraction of the exciton $|X|^2$ and photon $|C|^2$ in the polariton wavefunction, which is also called Hopfield coefficients (see Section 2 in SI).\cite{hopfield1958theory} The estimated coefficients as a function of the $k_x/k_0$ are shown in Figs. \ref{fig2}e-f.  At the $\Gamma$-point, where $k_x/k_0$ = 0, the exciton fraction $|X|^2$ is around 0.4 in the ASE sample and around 0.29 in the lasing sample. On the one hand the higher the exciton fraction, the larger the optical nonlinear effects and exciton optical gain, but on the other hand, with a higher exciton fraction, nonradiative losses rise, which reduces the polariton lifetime. As a result, there exists a balance between these two effects on the polariton branch, where polaritons may efficiently accumulate and cause polariton stimulation in this state after some threshold.\cite{shan2022brightening,zhang2022electric} 

\subsection{Polariton-mediated ASE and lasing in perovskite metasurface}

\begin{figure}[t!]
\centering
\center{\includegraphics[width=0.8\linewidth]{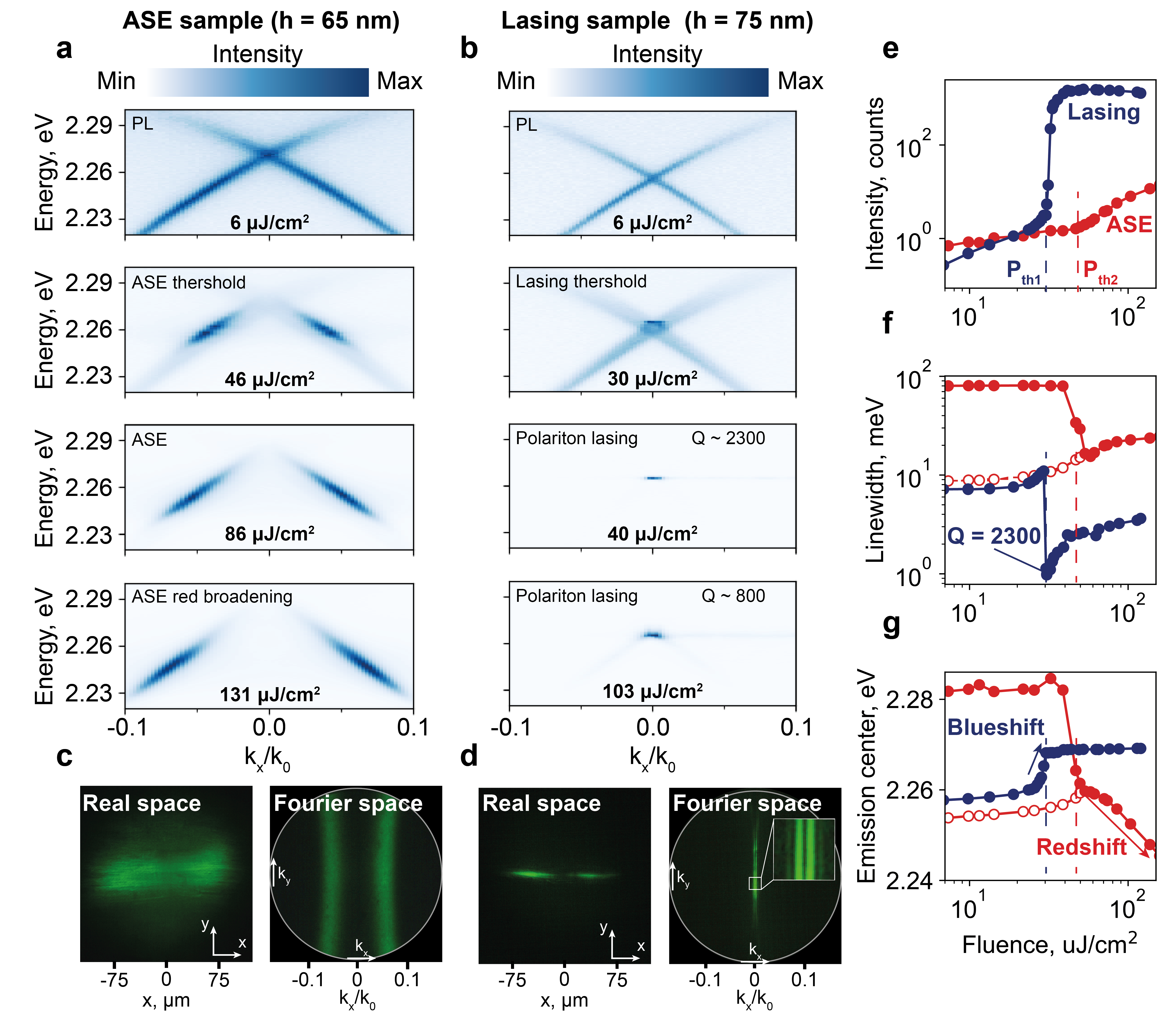}}
\caption{Polariton-mediated ASE and lasing in the samples without and with exceptional points, respectively. (a) Angle-resolved emission spectra obtained from ASE sample under different pump fluences. Linear PL is observed under 6~$\mu$J/cm$^2$, near the 46~$\mu$J/cm$^2$ intense broadband ASE peak appears, which is localized in the polariton branch and is of the polariton nature. With increasing the pump fluence, the ASE peak rises in intensity and broadens towards the red region due to the polariton relaxation. (b) Similar measurements are provided for the lasing sample. Under the pump fluence threshold of 30~$\mu$J/cm$^2$, narrow lasing emission appears around $k_x/k_0$ = 0, which rises in intensity and broadens with increasing pump fluence. The origin of the lasing in the polariton accumulation in the EPs (c) The images of the ASE in the real and Fourier space, obtained under the 86~$\mu$J/cm$^2$. Fourier space shows high-intensive isofrequency at the ASE spectral energies. (d) The images of the lasing emission, which are obtained under the 40~$\mu$J/cm$^2$ of pump fluence. The sample shows narrow lasing emission in \red{the} real space and narrow plane emission in the Fourier space along $k_x$. The characteristic dip in intensity at the $k_x$ = 0 is supposed to be due to the symmetry mismatching between the BIC and the plane wave propagating normally away from the sample. (e) \red{The} intensity of the emission as a function of the pump fluence. Solid red and blue lines with solid markers show the intensity of the emission observed in the ASE and lasing samples, respectively. (f) Linewidths of the ASE and lasing emission peaks as a function of the pump fluence. Solid red markers correspond to the linewidth of the integrated emission from the ASE sample over the $k_x/k_0$. Empty red markers correspond to the mode linewidths of the ASE sample at $k_x/k_0$ = 0.029. Solid blue markers correspond to the mode linewidth of the lasing sample at $k_x/k_0$ = 0. (g) Extracted emission central energy of the integrated ASE sample, mode of the ASE sample, and mode of the lasing sample, shown as red solid markers, red empty markers, and blue solid markers, respectively. 
}
\label{fig3}
\end{figure}

To study the stimulated emission of the samples, we perform angle-resolved PL measurements by varying the pump fluence. Under the pump fluences around 6~$\mu$J/cm$^2$ and lower we observe linear emission regime, \red{coming} from the polariton branches and background PL, which is observed from the unstructured film (Figs. \ref{fig3}a and 3b). The intensity of the background PL spectrum is much lower than the emission intensity of polariton modes and it is hidden in figures, however, it is still noticeable (see Fig. S7 in SI for details). With increasing the pump fluence, we observe a blueshift and broadening of the polariton modes, and after the pump threshold, we observe stimulated emission. In the case of the ASE sample, broad enhanced amplified spontaneous emission (ASE) on the polariton branch for $k_x/k_0$ in the range of 0.03-0.05 is observed, where the exciton Hopfield coefficient is around 0.3. But in the case of the lasing sample, we observe spectrally narrow emission with $k_x/k_0 \approx$~0, corresponding to the polariton accumulation in EPs. Since the metasurface with a height of $h =$~65~nm provides the spectral position of modes crossing closer to the exciton resonance in comparison with $h = $~75~nm, the non-radiative optical losses are higher (see SI for details). Also, the radiative losses are increased due to the higher relative modulation (see SI for details). Therefore, the full losses of the optical resonance at $k_x/k_0 = 0$ cannot be compensated with optical gain to achieve EPs and hence provide only ASE.

With increasing the pump fluence ASE shifts to the red region along the polariton branch and spectrally \red{broadens} (Fig. \ref{fig3}a). Meanwhile, the lasing emission increases in intensity, broads, and slightly shift to the blue region (Fig. \ref{fig3}b). The reason for the observed effect is supposed to be in the origin of the polariton relaxation.\cite{shan2022brightening,zhang2019photonics} In the case of the ASE, \red{the} accumulated polaritons in a nonlinear regime scatter to the lower energies, as there does not exist a state where a large number of polaritons can \red{be accumulated}. However, in the lasing sample, in a nonlinear regime, there appear the specific states at the $k_x/k_0 \approx$ 0, where the polariton condensation with a narrow lasing emission is achieved. As was mentioned above, the states are supposed to be the EPs, which are caused by the stimulated emission and will be described in detail below.

In \red{the} real space (Fig. \ref{fig3}c), the ASE emission is characterized as high-intensive spots with speckles on the left and right sides from the pump spot center with a wide distribution over the y-axis. Meanwhile, the lasing emission has a narrow distribution over the y-axis (Fig. \ref{fig3}d). In the Fourier space, ASE is almost uniformly distributed over the polariton branches, showing isofrequency contours in the range of the ASE spectral center (Fig. S6 in SI). However, the lasing emission is localized along the $k_x/k_0$ = 0 with almost a straight line with respect to the $k_y/k_0$ (see inset in Fig.\ref{fig3}d). As was shown before, the spectral position of the modes intersection is considered as a constant for the $|k_y/k_0| < 0.2$ (Fig.~\ref{fig1}e), and therefore EPs are supposed to exist over the whole of this stripe. As a result, polaritons accumulate equiprobably over the stripe and produce the unidirectional lasing. It should be noted that in the Fourier space image of the lasing, we detect a characteristic dip in the lasing intensity exactly around $k_x/k_0$ = 0 because of the symmetry mismatching between the BIC and plane wave propagating normally. The EPs \red{appearing} in the nonlinear regime are slightly detuned from the $k_x/k_0$ = 0. It is not observed in the spectra of the lasing emission, shown in Fig. \ref{fig3}b, because of the limited resolution over\textit{ $k_x/k_0$}, and will be shown further. 

In order to analyze the ASE and lasing emission as a function of the pump fluence in detail, we fit it by the Lorentzian function. As ASE is not localized in some $k_x/k_0$, we integrate the emission overall $k_x/k_0$ for the analysis (solid red markers in Figs. \ref{fig3}e-g), and for the lasing sample, we study the emission at the $k_x/k_0$ = 0 (solid blue markers in Figs. \ref{fig3}e-g). However, in order to compare the mode parameters of the ASE sample before the threshold, we also fit the mode at $k_x/k_0 \approx$ 0.03, corresponding to the maximum of the ASE at the pump threshold (empty red markers in Figs.~\ref{fig3}f-g). Both ASE and lasing emission show the S-curve \red{shape of dependence of the emitted signal} intensity as a function of the pump fluence (Fig. \ref{fig3}e). The pump threshold of the ASE is estimated as 46~$\mu$J/cm$^2$ and of the lasing emission as 30~$\mu$J/cm$^2$. The difference in the pump threshold is attributed to the different losses of the samples, mentioned above and also \red{the} existence of the EPs in the lasing sample. The EPs appear in the system in a nonlinear regime when polariton relaxation provides the optical gain and does not require overcoming all losses (see Section 3 of SI). When it appears, the probability of the polaritons scattering in this state dramatically increases, which reduces the lasing threshold. Also, the lasing sample shows an enhanced increase in the intensity after the threshold by more than three orders of magnitude in comparison with the one order observed in the ASE sample. This can be explained by the dramatically increased rate of polariton stimulation to EPs in \red{the} nonlinear regime.\cite{zhang2022electric} \red{The} polaritons \red{reaching} this state radiatively recombines, producing lasing emission in a time scale much shorter than a non-radiative lifetime,\cite{schlaus2019lasing} which thus strongly suppresses non-radiative recombination. \red{The slight decrease of the lasing intensity at higher fluence can appear due to the further increase of the optical gain, leading to the disappearance of the polariton mode degeneracy and the fade of the exceptional points. Also, it can be attributed to Auger recombination or additional heating of the material. \cite{lei2021metal}}

The integrated emission of the ASE sample in the linear regime before the threshold represents \red{a} broad spectrum with a linewidth around 80~meV and a center around 2.28~eV, as shown in Figs. \ref{fig3}f and \ref{fig3}g by red solid markers. It should be noted that the estimated linewidth here is broadened with respect to the background PL spectrum, because \red{the} leaky modes enhance the outcoupled emission in the red region (see Fig. S7 in SI for details). At the threshold, the linewidth rapidly \red{drops} to the values of the mode linewidth (shown as red empty markers), as well as the emission center, because the ASE is strongly localized and enhanced by the polariton mode as it is shown in Fig.~\ref{fig3}a. It should also be noted that exactly before the threshold the center of the mode slightly shifts to the blue region, which is attributed to the nonlinear polariton blueshift under \red{the} non-resonant pump.\cite{ardizzone2022polariton,deng2010exciton} With increasing the pump fluence, ASE \red{spectrum} broadens and shifts to the red region, because of the polariton relaxation discussed above. The lasing sample \red{exhibits} the mode linewidth of around 6~meV before the threshold with a spectral center energy of~2.258~eV. With further elevating the pump fluence, the mode shifts to the blue region by a value of 10~meV and broadens due to the polariton-polariton interaction.~\cite{masharin2022polaron,masharin2022room,deng2010exciton} Around the threshold, narrow lasing emission appears with a linewidth of around 1~meV corresponding to a Q-factor of 2300. With further increasing the pump fluence, the lasing mode broadens and slightly shifts to the blue region. The broadening is supposed to be due to the temporal changing of the lasing center peak during the emission. The blueshift of the lasing peak is determined by the number of polaritons. After the high-intensity pump pulse is absorbed, the formed polaritons shift the polariton branch to the blue region and the lasing emission appears. When the polaritons annihilate, producing the lasing emission, the number of polaritons reduces and the branch starts to return back to its initial spectral position. As a result, we observe broadened integrated-over-time lasing spectra. This assumption was previously confirmed by the temporal resolution of the lasing emission in polariton systems with a streak-camera.\cite{ardizzone2022polariton,schlaus2019lasing}

\subsection*{Observation of the EPs in the lasing regime}

\begin{figure}[t!]
\centering
\center{\includegraphics[width=0.8\linewidth]{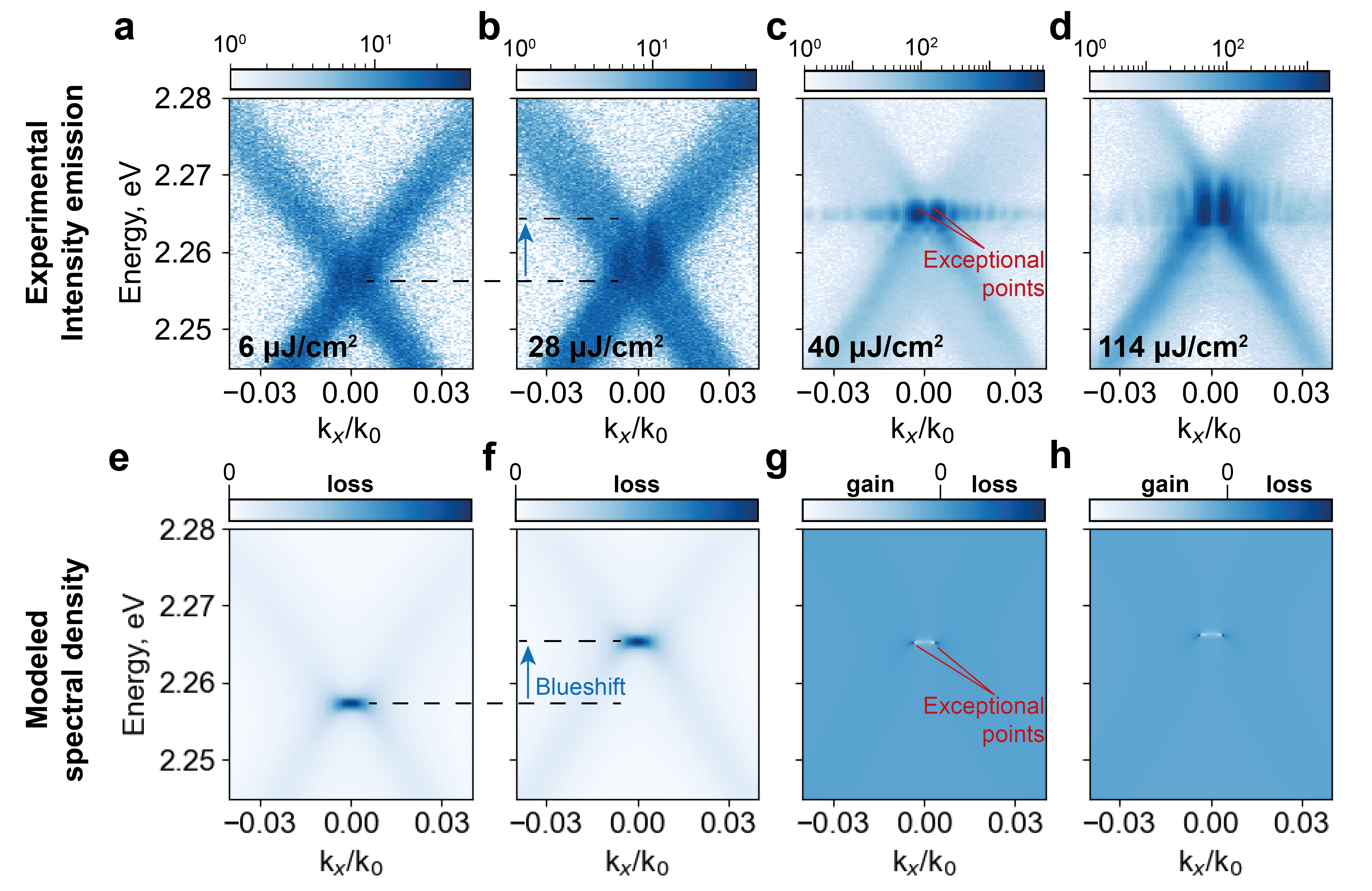}}
\caption{The observation of exceptional points in the lasing regime. 
(a-d) Measured angle-resolved spectra of the lasing sample, obtained with enhanced resolution over a $k_x/k_0$ axis at $k_y/k_0$ = 0. In the linear PL spectrum, obtained under pump fluence lower than 6~$\mu$J/cm$^2$, we observe enhanced spontaneous emission near the $k_x/k_0$ = 0, which is attributed to the EPs.\cite{lu2020engineering,ferrier2022unveiling} With the increase of pump fluence, we observe two lasing spots and a dark spot at $k_x/k_0$ = 0, which was barely visible in Fourier image in Fig \ref{fig3}b, which rises in intensity with the increase of the pump fluence. Under 114~$\mu$J/cm$^2$ we observe the polariton lasing blueshift, due to the polariton-polariton interaction, and also spectral broadening discussed in the text in detail. (e-h) Modeled spectral density of the studied perovskite metasurface lasing sample corresponding to the following regimes: (e) linear regime in the absence of optical gain, originating from the polariton stimulation, where the simulated quantity corresponds to the emission, observed in the experiment; (f) linear regime with blueshifted polariton branches due to the polariton-polariton interaction under pump fluences around 28~$\mu$J/cm$^2$; (g) stimulated regime calculated by subtracting the gain profile (see Section 3 of SI), where the change of the sign of the spectral density corresponds to the onset of the EPs providing enhanced LDOS; (h) corresponds to the system with an enhanced gain profile and slightly shifted polariton branch, as it is observed in the experiment at the highest fluence values.}
\label{fig4}
\end{figure}

To confirm the presence of the EPs in the stimulated emission of the lasing sample, we increase the magnification of the BFP image, \red{and} transferred to the spectrometer imaging CCD to increase the angular resolution over $k_x/k_0$. The angle-resolved spectra with \red{the} increased resolution measured for different pump fluences with logarithmic scale in the pseudocolor are shown in Fig.~\ref{fig4}a. Under the low pump fluence around 6~$\mu$J/cm$^2$, we observe a slight enhancement of the spontaneous emission near the $k_x/k_0$ = 0, which can be caused by the enhanced spontaneous emission, also reported in ref. \cite{ferrier2022unveiling}. At the pump threshold around 30~$\mu$J/cm$^2$, polaritons start to interact with each other, which leads to the polariton branch blueshift and polariton stimulated scattering to the lower energies. The latter causes the polariton accumulation in the spectral region of observed ASE and provides the optical gain for leaky modes. As a result, when the critical values of the optical gain are achieved, EPs appear, and we observe two distinguished lasing spots near the $k_x/k_0$ = 0 (Fig. \ref{fig4}c). Thanks to the enhanced LDOS, polaritons occupy these states, which leads to the polariton lasing. With increasing the pump fluence, we observe the increase of the lasing peak intensity and then broadening and the blueshift mentioned above (Figs. \ref{fig4}c-d). In the logarithmic scale scattered lasing signals over all $k_x/k_0$ are observed, which are attributed to the scattering on the sample impurities and grain boundaries. Also, \red{around fluence} 114~$\mu$J/cm$^2$ it can be observed that \red{the} broadened and blueshifted emission preserves the radiation \red{pattern}, which is related to the idea of temporal lasing peak broadening discussed above (Fig. \ref{fig4}d).

To model \red{the} observed phenomena of EPs we calculate the eigenvalues and spectral density of the optical modes, which can be estimated as an imaginary part of Green's function, $S(\omega,k_x) = Im(\hat{G}(\omega,k_x))$ identified as 
$\hat{G}(\omega,k_x) = (\omega-\hat{H}(k_x))^{-1}$, where $\omega$ is the frequency and $\hat{H}(k_x)$ is the Hamiltonian of the metasurface cavity modes and can be written as: \cite{engdahl1986generalized, vijay2002polynomial, lu2020engineering}

\begin{equation}
    \hat{H}(k_x) = 
    \begin{pmatrix}
    E_+(k_x) & U\\
    U & E_-(k_x)
    \end{pmatrix}
    - i
    \begin{pmatrix}
    \red{\gamma}_{nr}(k_x) + \red{\gamma}_r & \red{\gamma}_r \\
     \red{\gamma}_r & \red{\gamma}_{nr}(k_x) + \red{\gamma}_r
    \end{pmatrix},    
\end{equation}
where $E_{+,-}(k_x)$ stands for the real part of the leaky mode dispersion with positive and negative group velocity, respectively, $\red{\gamma}_{nr,r}$ stands for the non-radiative and radiative losses, respectively, and $U$ stands for the coupling coefficients between modes. We extract mode centers and linewidths from the experimental data, shown before. As we do not observe the splitting between two modes at $k_x/k_0$ = 0, we take parameter $U$ equal to zero. EPs appear in the system when the eigenvalues of the Hamiltonian coalesce:

\begin{equation}
    \hat{H} \mathbf{a} = \lambda \mathbf{a},
\end{equation}

\begin{equation}
    \lambda_{1}(k_x^{EP}) = \lambda_{2}(k_x^{EP}),
\end{equation}

In this case, spectral density $S(\omega^{EP},k_x^{EP})$ possess artifacts, which are observed in the experiment as lasing points.

The result of the spectral density calculation based on the extracted experimental data in linear regime is shown in Fig. \ref{fig4}e. The "hotspot" at the \red{$\Gamma$-point}, where $k_x/k_0$ = 0 corresponds to the enhanced spontaneous emission, shown in Fig.~\ref{fig4}a. Then, we shift the polariton mode in the model as it is observed in the experiment (Fig. \ref{fig4}f). To take into account the stimulated emission, we assume that \red{the} nonradiative losses can be expressed as the difference between absorption (losses) $\alpha(k_x)$ and gain (stimulated polaritons) $\beta(k_x)$: $\red{\gamma}_{nr}(k_x) = \alpha(k_x) - \beta(k_x)$. To estimate the $\beta(k_x)$ we extract the ASE spectrum from the experimental results, shown in Fig.~\ref{fig3}a, and consider it as a gain $\beta(k_x)$, multiplied by the constant $G_0$. By varying the parameter $G_0$ we find the critical conditions when the eigenvalues degenerate and EPs appear (see Section 3 of SI for details). After \red{the} applying the calculated gain, which does not exceed the full losses, we observe the sign changing of spectral density in the particular places near $k_x/k_0$, which corresponds to the EPs. As EPs are the  specific state, which is very sensitive to the variation of the parameters,\cite{miri2019exceptional} it has to be destroyed after the higher $\beta(k_x)$ under higher fluences. However, thanks to the polariton nature, in the regime of the polariton condensation in EPs, all new polaritons which scatter under a higher pump rapidly recombine and do not produce the change of the imaginary part of the mode.

\section{Conclusion}

In summary, we have demonstrated room-temperature polariton lasing mediated by the EPs in the perovskite-based non-local metasurface. Along with the solution-processable perovskite synthesis methods, the nanoimprint lithography provides large-area gratings supporting high-Q modes (up to Q$\approx$2300) formed by symmetry-protected BIC and the EPs in the active nonlinear regime. We have revealed that the ASE in the studied samples originates from the accumulation of polaritons in the spectral region with the excitonic fraction in polariton of around 0.3. With further increase of the carrier concentration, the ASE broadens towards the red region due to the polariton relaxation, and EPs appear when the metasurface modes are crossed within the gain spectral range.
Our results explain the fundamental origin of the lasing in lead-bromide metasurfaces and open new ways for the realization of room-temperature planar polariton lasers based on a rich variety of halide perovskites. Since the proposed cavity design is close to the most optimal case, we envision further lasing thresholds lowering through the material properties optimization via \red{the} synthesis of highly crystalline perovskite films~\cite{du2022all, Pourdavoud2018Apr, tatarinov2023high} and cation engineering~\cite{alvarado2022lasing}, \red{which can lead to the low-cost continuous wave~\cite{jia2017continuous, qin2020stable} and electrically pumped lasers~\cite{wang2022electrically}, essential for industrial applications.}

\section{Experimental Section}
\threesubsection{Perovskite metasurface fabrication}\\
The fabrication process is divided into three main steps: solution preparation, thin film spin-coating, and nanoimprint lithography. First, the perovskite solution is prepared in the dry N$_2$ glove box by the mixture of 33.59 mg of MABr (TCI) and 110.1~mg of PbBr$_2$ (TCI). Then it is dissolved in a 1 mL mixture of DMF and DMSO in a ratio of 3:1 and steered by \red{a} magnetic stirrer for 24 h. Second, substrates of SiO$_2$ are cleaned by consistent sonication in deionized water, acetone, and isopropanol for 10 min. After it is dried out and put into the plasma cleaner for 10 min to achieve high surface adhesion. Cleaned substrates are transferred to the dry N$_2$ glove box, where further film deposition is provided. The perovskite film synthesis starts with the deposition of 30~$\mu$L MAPbBr$_3$ solution on the cleaned substrate and then the spin-coater starts the rotation. The substrate with a deposited solution is spun for 50 s with a speed rotation of 3.000 or 3.500 rates per minute (rpm), depending on the target film thickness. At the 35th s, after the rotation starts, 300~$\mu$L of the antisolvent (toluene) is dripped at the top of the rotated substrate to remove the solution and cause fast and uniform film crystallization. \red{The surface morphology, XRD patterns, and absorbance spectrum are in good agreement with previous works \cite{Wang2021Jun, soufiani2015polaronic} and are shown in Figure S11, S12, and S13, respectively, in SI.} At the third stage of the synthesis substrate with perovskite film in the intermediate phase before thermal annealing is transferred to a laboratory press for the nanoimprint lithography. The mold for the nanoimprint represents the periodic grating with a period of 320~nm, a comb height of 25~nm, and a ratio between comb width and structure period (FF) of 0.5, made of polycarbonate. The perovskite film with the mold on the top is located in the laboratory press, where the pressure of around 200 MPa is applied for 10 min. After the mold is removed, the substrate with the perovskite imprinted film is annealed at 90~$^\circ$C for 10~min in the atmosphere of the dry N$_2$ glove box.

\threesubsection{Angle-resolved spectroscopy}\\
 Angle-resolved spectroscopy measurements are performed with a back-focal-plane (BFP) 4f setup by using a slit spectrometer, coupled to the imaging EMCCD camera (Andor Technologies Kymera 328i-B1 + Newton EMCCD DU970P-BVF) and a halogen lamp, coupled to the optical fiber with the collimation lens, employed for the white light illumination. Olympus Plan Achromat Objective with a magnification of 10x and numerical aperture of 0.25 is used for the excitation and collection of the signal. Spatial filtering is carried out in the intermediate image plane (IP), transferred by the 4f scheme. The polarization is filtered by a linear polarizer after the spatial filter. Angle-resolved photoluminescence measurements are performed in the same setup by using the femtosecond (fs) laser with a wavelength of 400~nm and a repetition  rate of 1 kHz. The laser system represents mode-locked Ti:sapphire laser at 800~nm, which is used as a seed pulse for the regenerative amplifier (Spectra Physics, Spitfire Pro), and then is frequency-doubled via BBO crystal. A lens with a focal distance of 500~mm is used to focus the laser beam in the BFP to achieve a large pump spot of around 175~$\mu$m. To filter the laser in the collection channel long-pass filter FEL450 is used. A CCD camera (Thorlabs 1.6 MP Color CMOS Camera DCC1645C) with a 150~mm tube lens after the beamsplitter in the collection channel is used for the imaging of the real and Fourier space. The experimental scheme is shown in Fig. S1 in SI.

\medskip
\textbf{Supporting Information} \par 
Supporting Information is available from the Wiley Online Library or from the author.

\medskip
\textbf{Acknowledgements} \par 
The authors thank Furkan Isik for the provided SEM pictures \red{and absorbance spectra }of the samples. \red{The authors also thank Farzan Shabani for the provided XRD pattern measurements.} The work was partially done in ITMO Core Facility Center "Nanotechnologies". A.K.S. acknowledges the Deutsche Forschungsgemeinschaft (Grant SFB TRR142/project A6), the Mercur Foundation (Grant Pe-2019–0022), and TU Dortmund core funds. This work was also supported by the Ministry of Science and Higher Education of the Russian Federation (Project 075-15-2021-589) and Priority 2030 Federal Academic Leadership Program. H.V.D. acknowledges the support from JUBA.

\medskip

%

\bibliographystyle{MSP}
\bibliography{main}

\include{Supplementary}

\end{document}

%% file: Supplementary.tex
\renewcommand{\thefigure}{S\arabic{figure}}


\title{Supplementary Information: Room-temperature exceptional-point-driven polariton lasing from perovskite metasurface}

\maketitle


\author{M.A.~Masharin}
\author{A.K.~Samusev}
\author{A.A.~Bogdanov}
\author{I.V.~Iorsh}
\author{H.V.~Demir}
\author{S.V.~Makarov*}


\dedication{}

\begin{affiliations}
M.A.~Masharin, H.V.~Demir\\
Institute of Materials Science and Nanotechnology and National Nanotechnology Research Center and Electronics Engineering and National Nanotechnology Research Center, Department of Electrical, Department of Physics, Bilkent University, Ankara, 06800, Turkey\\

M.A.~Masharin, A.K.~Samusev, A.A.~Bogdanov, I.V.~Iorsh, S.V.~Makarov\\
ITMO University, School of Physics and Engineering, St. Petersburg, 197101, Russia\\
Email Address: s.makarov@metalab.ifmo.ru\\

A.K.~Samusev\\
Experimentelle Physik 2, Technische Universit\"at Dortmund, 44227 Dortmund, Germany\\

A.A.~Bogdanov, S.V.~Makarov\\
Qingdao Innovation and Development Center, Harbin Engineering University, Qingdao 266000, Shandong, China\\

I.V.~Iorsh\\
Department of Physics, Bar-Ilan University, Ramat Gan, 52900, Israel\\

H.V.~Demir\\
LUMINOUS! Center of Excellence for Semiconductor Lighting and Displays, School of Electrical and Electronic Engineering, School of Physical and Materials Sciences, School of Materials Science and Engineering, Nanyang Technological University, 639798, Singapore\\

\end{affiliations}


\keywords{Exciton-polariton condensation, polariton lasing, perovskite metasurface, exceptional points}

\section*{Calculation of the perovskite PCS isofreqencies and the dispersion surface.}

\begin{figure}[t!]
\centering
\center{\includegraphics[width=0.8\linewidth]{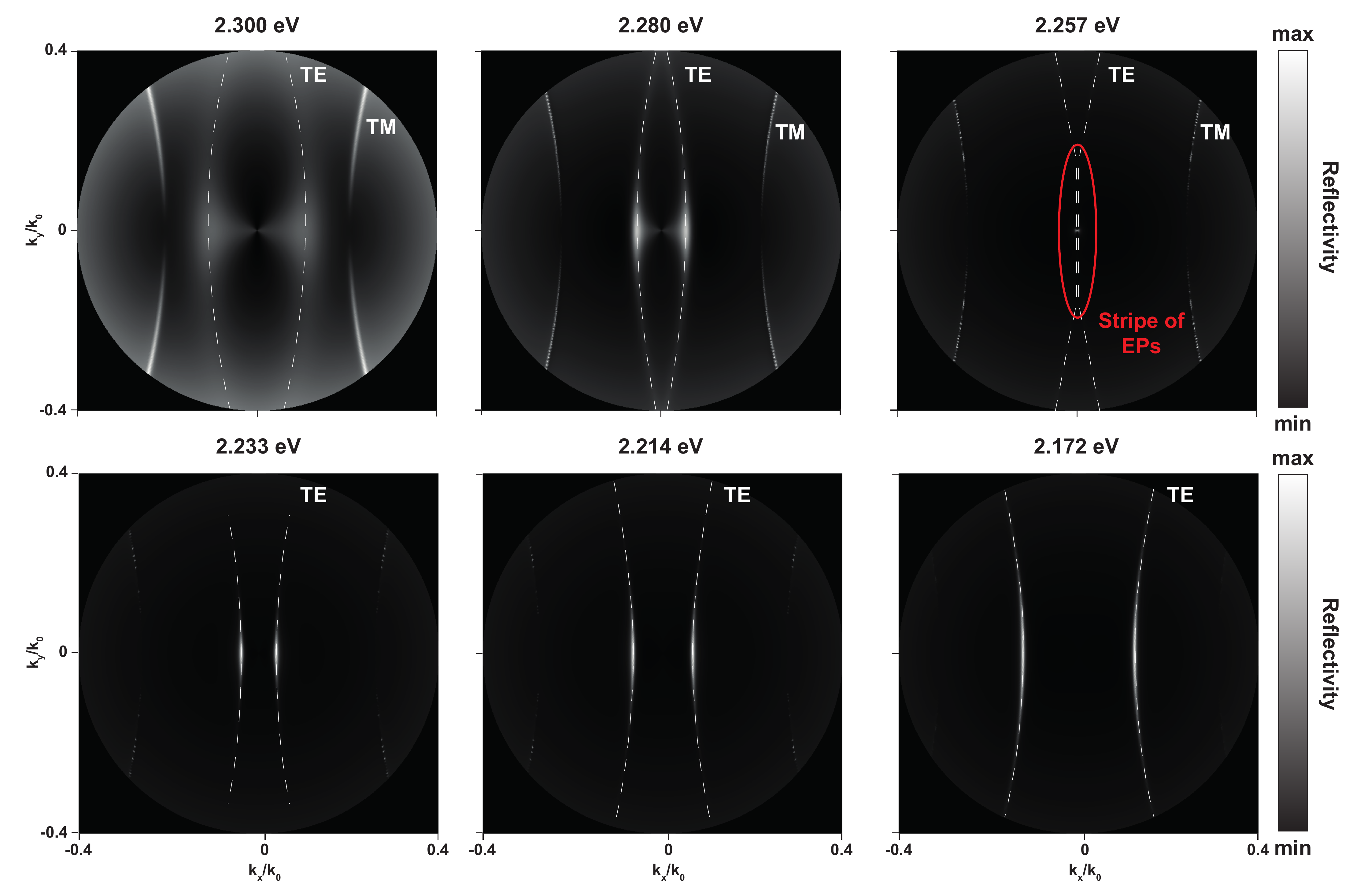}}
\renewcommand\thefigure{S6}
\caption{Isofrequences of the studied perovskite PCS, for different energies (frequencies). Dashed white lines correspond to the extracted modes. On the plots TE and TM modes are shown, which correspond to the leaky modes with an electric field and magnetic, which are co-directed with the gratings respectively. The red ellipse shows the stripe of EPs, observed at the resonant energy.}
\label{figSisofreq}
\end{figure}

In order to plot the dispersion surface, we calculate isofrequencies of perovskite PCS at different frequencies (energies), shown in Fig. \ref{figSisofreq}. In the range of wavenumbers |$k_x/k_0$| and |$k_y/k_0$| < 0.4 we observe two modes, corresponding to the TE and TM polarizations. TE mode has an electric field codirect with the PCS combs, and TM has a magnetic field co-directed with this direction. In the work, we study only TE mode, as it has stronger electric field localization in the material and therefore is strongly coupled to the exciton resonance. Moreover, in this geometry, only the TE mode has the EPs in the region of optical polariton gain. However, it is possible to achieve EPs in TM geometry with the variation of the sample thickness. We extract the isofreqency curves of TE mode at each energy and assemble them in the one dispersion surface, shown in the main text. It should be noted, that the stripe of the intersection points, which are identified as EPs in the experiment, is observed in calculated FMM isofrequency at 2.257 eV. 

\begin{figure}[t!]
\centering
\center{\includegraphics[width=1\linewidth]{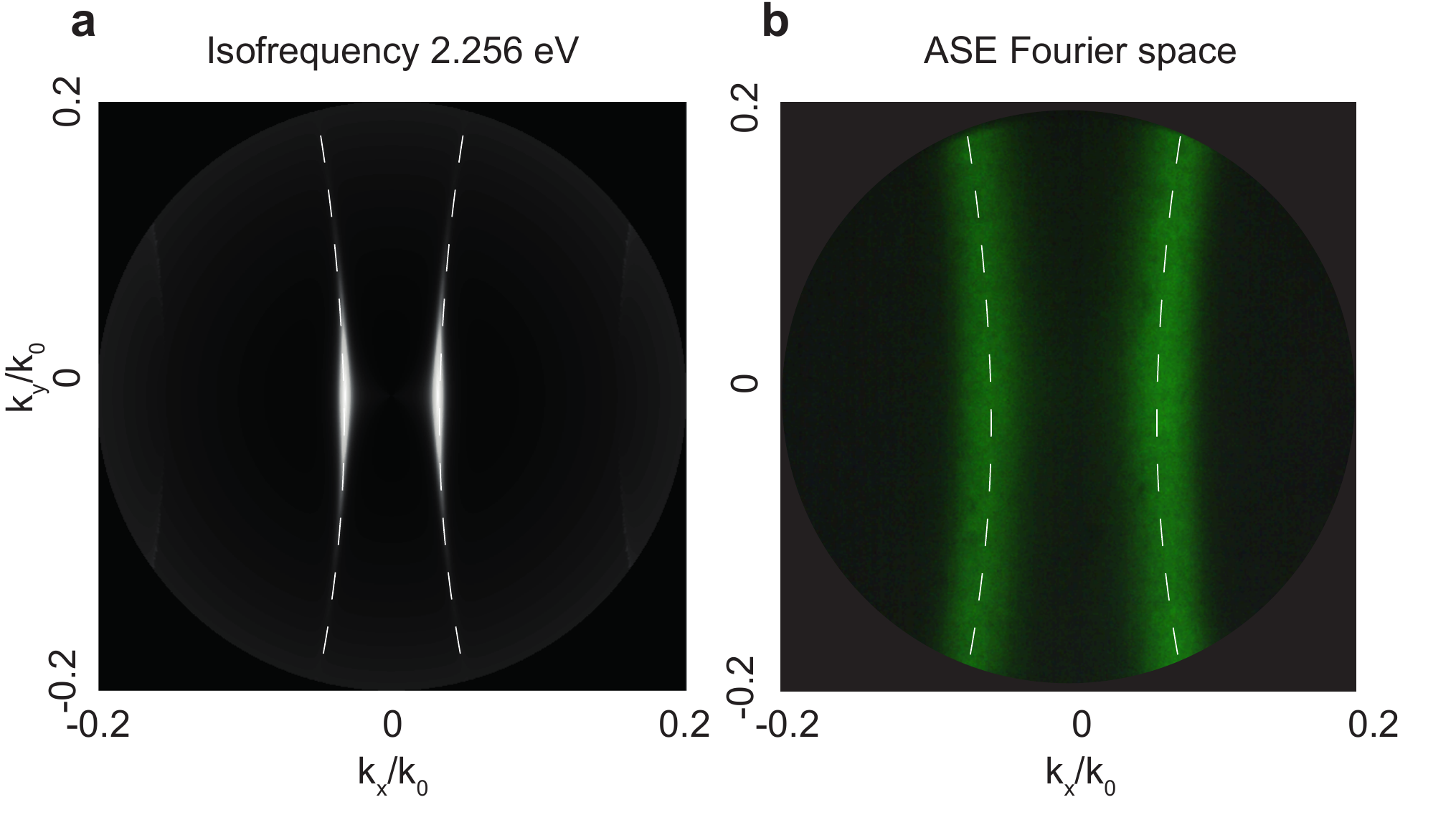}}
\renewcommand\thefigure{S7}
\caption{(a) Calculated isofreqency of the ASE sample at the 2.256 eV (b) Measured Fourier space of the ASE under 49 $\mu$J/cm$^2$ of pump fluence. The dashed blue lines correspond to the estimated centers of the mode.}
\label{figSisofreqASE}
\end{figure}

The calculated isofrequency of the ASE sample in the spectral region of the ASE emission with measured Fourier plane is shown in Fig. \ref{figSisofreqASE}. It is shown, that in the experiment the ASE emission is strongly broadened in comparison with the calculation. It is because of the ASE broad spectrum, which contains a broad region of the wavenumbers. Nevertheless, calculated isofreqency in the ASE spectral center well correspond to the experimental observations.

\section*{Calculation of the Hopfield coefficients}

According to the two-coupled oscillator model the polariton state consists of the exciton and photon with their weights in the wavefunction, depending on the wavevector $k_x$. The exciton and photon fractions is labeled as $|X_k|$ and $|C_k|$ respectively and based on the coupling coefficient $g$ can be calculated as \cite{hopfield1958theory}

\begin{equation}
    |X_k|^2 = \frac{1}{2}\left(1 + \frac{|E_X-E_C|}{\sqrt{(E_X-E_C)^2 + g^2}}  \right),
    \\
    |C_k|^2 = \frac{1}{2}\left(1 - \frac{|E_X-E_C|}{\sqrt{(E_X-E_C)^2 + g^2}}  \right)
\end{equation}

\section*{Determination of the optical gain in the EPs model.}

\begin{figure}[t!]
\centering
\center{\includegraphics[width=1\linewidth]{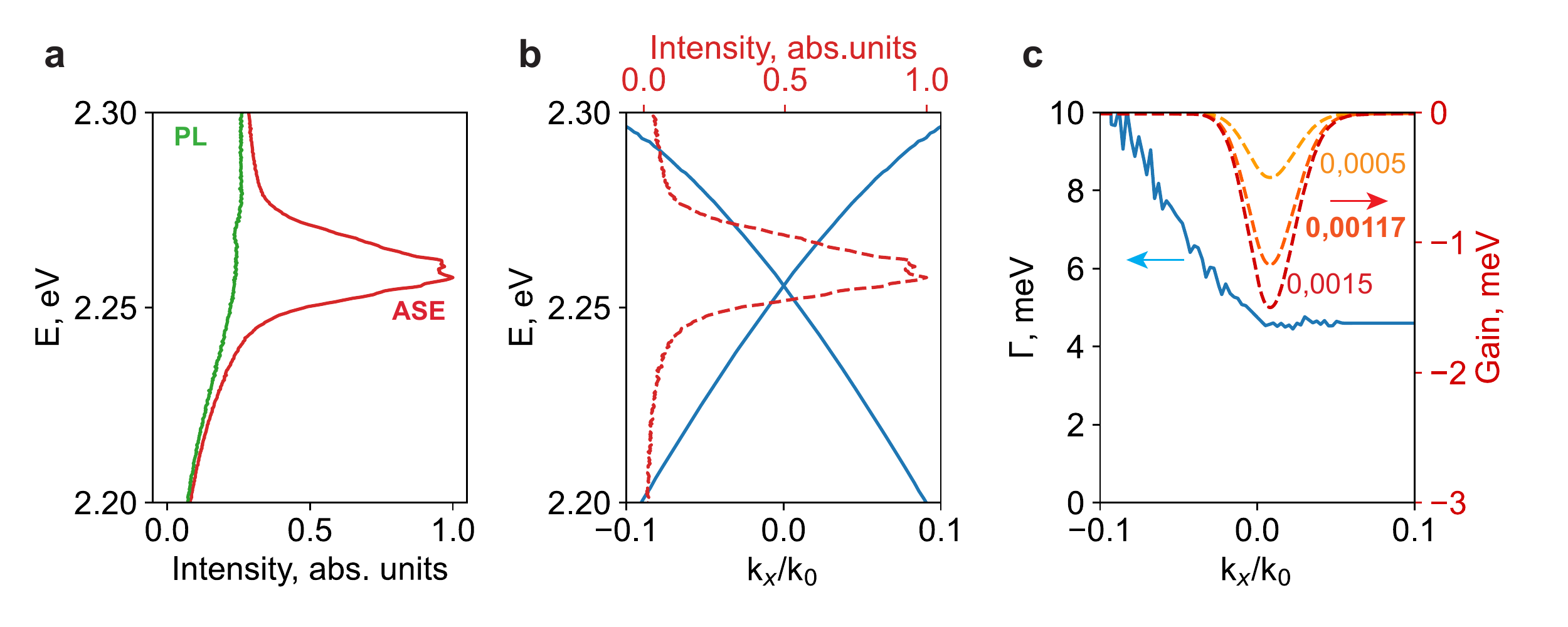}}
\renewcommand\thefigure{S9}
\caption{(a) Extracted integrated normalized PL spectrum, measured from the ASE sample under 6 $\mu$J/cm$^2$ of pump fluence, shown as a green solid line. Extracted normalized ASE emission, measured under 47 $\mu$J/cm$^2$ of pump fluence (b) Extracted modes of the studied lasing sample are shown as blue solid lines. The ASE spectrum extracted from the data in the previous panel is shown as a red dashed line. The figure shows the spectral region of the polariton optical gain of the ASE sample in the leaky modes dispersion of the lasing sample. (c) The linewidth of the lasing sample leaky mode with positive group velocity is shown as a blue solid line. The estimated ASE spectrum as a function of $k_x/k_0$ by the group velocity for different amplitude is shown as dashed orange lines. The values, shown in the figure correspond to the amplitude of the ASE spectra, used in the model for the EPs.}
\label{figSgainforModel}
\end{figure}

The optical gain in the polariton system appears due to the polariton accumulation in the particular state on the polariton branch, where there is an extremum between the scattering probability and the lifetime.\cite{zhang2022electric,shan2022brightening} However, calculating the optical gain is a challenging task, and therefore we roughly estimate it from the experimental results in order to show qualitatively the origin of EPs. As two of our samples: the ASE sample and the lasing sample are very similar in terms of the mode dispersions, we extract the ASE spectrum, measured from the ASE sample (Fig. \ref{figSgainforModel}a). The spectral region of the ASE is in the range of the leaky modes intersection of the lasing sample (Fig. \ref{figSgainforModel}b). We assume, that in the lasing sample the spectral region of the polariton accumulation can be considered the same as in the ASE sample. As the spectral position of the mode is connected with $k_x/k_0$ we calculate the gain profile in terms of the $k_x/k_0$, as $\beta(k_x) = G_0 \cdot ASE(k_x)$, where $ASE(k_x)$ is normalized ASE spectrum, extracted from the experimental data and $G_0$ is the amplitude of the gain.

As was described in the main text, we subtract gain profile $\beta(k_x)$ from the full optical losses: $\red{\gamma} = \alpha(k_x) - \beta(k_x)$, where $\alpha(k_x)$ is determined as linewidths, extracted from the PL, measured under 6 $\mu$J/cm$^2$ of pump fluence, shown in Fig.\ref{figSgainforModel}c.  

\begin{equation}
    \hat{H}(k_x) = 
    \begin{pmatrix}
    E_+(k_x) & U\\
    U & E_-(k_x)
    \end{pmatrix}
    - i
    \begin{pmatrix}
    \red{\gamma}_{nr}(k_x) +  \red{\gamma}_r &  \red{\gamma}_r \\
      \red{\gamma}_r &  \red{\gamma}_{nr}(k_x) +  \red{\gamma}_r
    \end{pmatrix}    
\end{equation}

With varying of the parameter $G_0$ we calculate real and imaginary parts of the hamiltonian eigenvalues as a function of $k_x/k_0$, shown as dashed blue lines in Fig. \ref{figSanalysisEPsAppear}. We put $U$ = 0, as we did not observe the splitting between two modes at the intersection point. To observe the appearance of the EPs we plot the eigenvalues difference of the real and imaginary parts ($\Delta E$, $\Delta \red{\gamma}$), shown in Fig.\ref{figSanalysisEPsAppear}. In the linear regime, where $G_0$ = 0 we observe the splitting between eigenvalues (Fig. \ref{figSanalysisEPsAppear}a). With increasing $G_0$ we observe that the difference between the real and imaginary parts of eigenvalues starts to decrease (Fig. \ref{figSanalysisEPsAppear}b). At the critical value of $G_0$, equal to 0.00117, we observe the degeneracy of the eigenvalues, at some particular values of $k_x/k_0$, which are supposed to be the appearance of the EPs (Fig. \ref{figSanalysisEPsAppear}c). Also, EPs are considered a very sensitive state to the environment, and as we also observe, if we increase $G_0$ more, the degeneracy disappears (Fig. \ref{figSanalysisEPsAppear}d). 

However, due to the nature of the polariton lasing, we assume, that in the experiment the gain never exceeds the critical value. When the gain achieves the critical value, the EPs polariton condensation state appears there. It means, that all new polaritons in the state promptly recombine, emitting coherent photons. In this polariton system, as described in this work, EPs can be achieved through nonlinear processes of the polariton relaxation, which results in the polariton lasing.

\begin{figure}[t!]
\centering
\center{\includegraphics[width=1\linewidth]{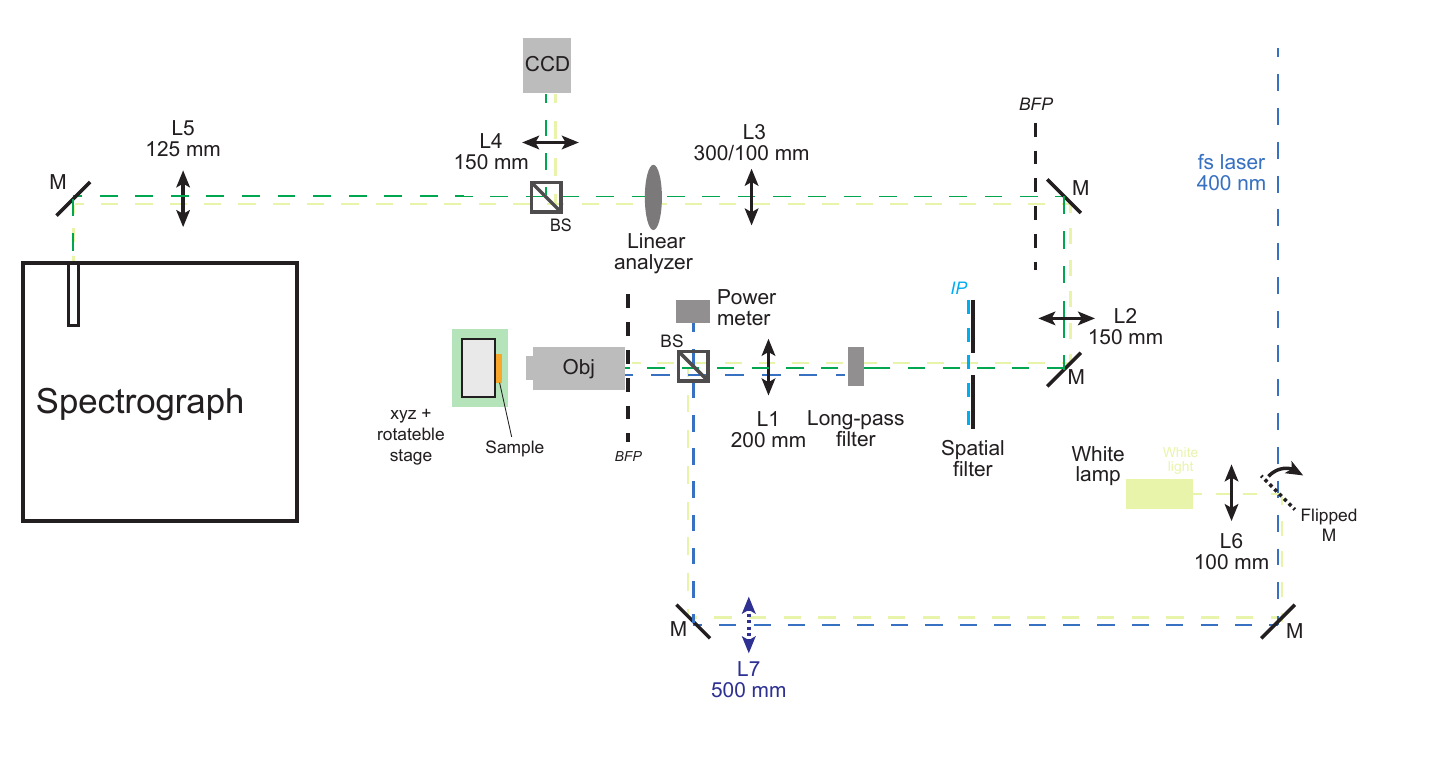}}
\renewcommand\thefigure{S1}
\caption{The scheme of the optical experimental setup. For the angle-resolved spectroscopy 4-f scheme is realized with lenses L1-L3. M stands for silver mirrors. IP stands for the transferred image plane, where spatial filtering is realized. BFP stands at the back focal plane. Linear analyzer filter TE linear polarization. L5 lens is used for the focusing of the signal to the spectrograph slit. The Yellow dashed line corresponds to the white light, which is collimated by the L6 lens. L7 is used for the focusing of the laser spot to the BFP of the objective. Blue dashed lines correspond to the laser pump pulse, which is filtered in the collection channel by the Long-pass 450 nm filter. BS stands for the beamsplitter Power meter, placed after the first BS is used for the pump fluence control. CCD camera with the tube lens L4 is used for real and Fourier space imaging. L3 is the removable lens, which makes the Fourier transform. By the substitution of 300 mm focal distance to 100 mm we increased the BFP for better resolution over $k_x/k_0$ in the experiment.}
\label{figSsetup}
\end{figure}

\begin{figure}[t!]
\centering
\center{\includegraphics[width=0.8\linewidth]{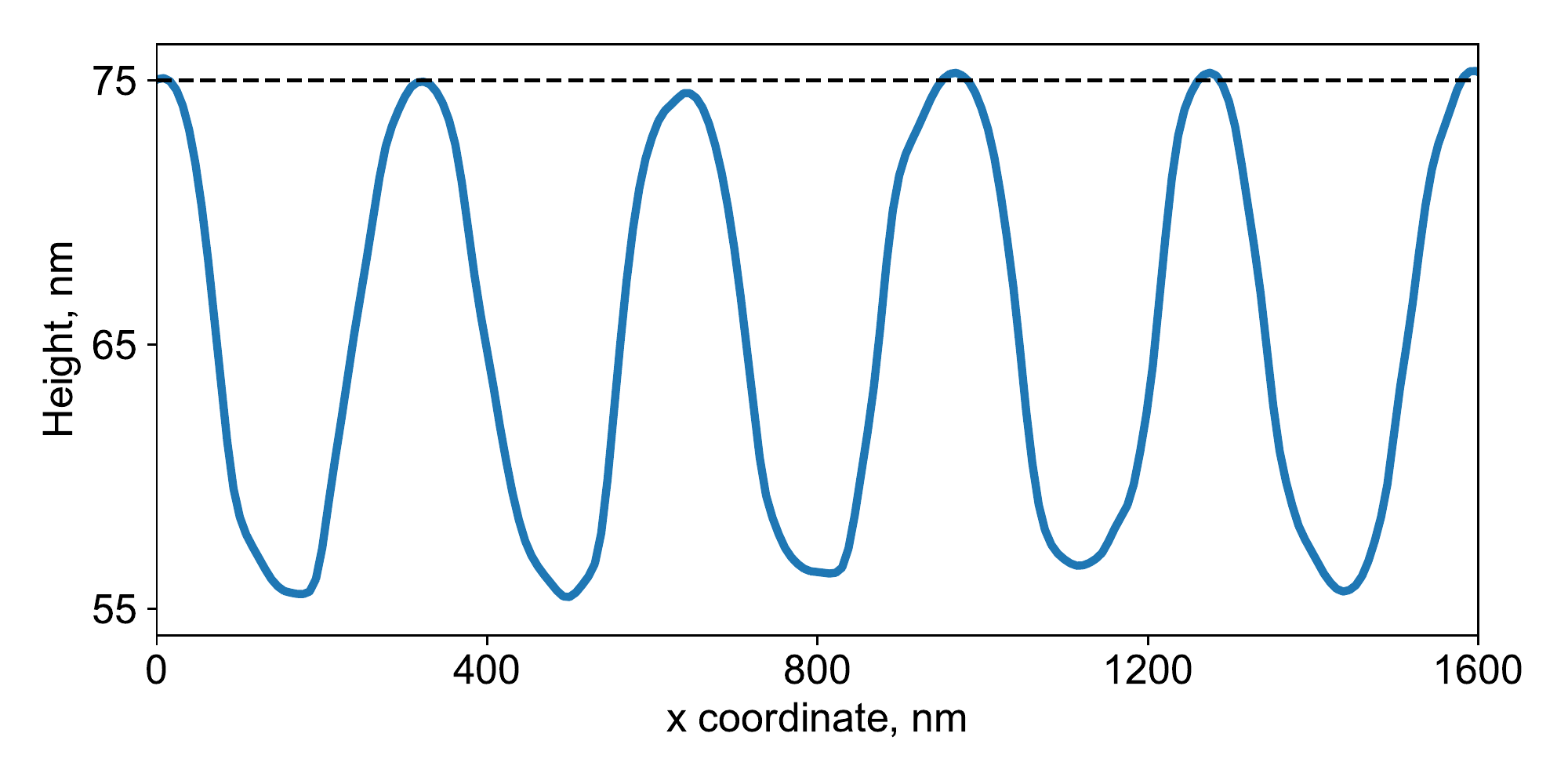}}
\renewcommand\thefigure{S2}
\caption{The profile of the studied PCS, extracted from the AFM measurements of the sample with the height of 75 nm. The comb height is around 20 nm, the period of the structure is 320 nm and the comb width, calculated as a FWHM of the comb is around 160 nm.}
\label{figSAFM}
\end{figure}

\begin{figure}[t!]
\centering
\center{\includegraphics[width=1\linewidth]{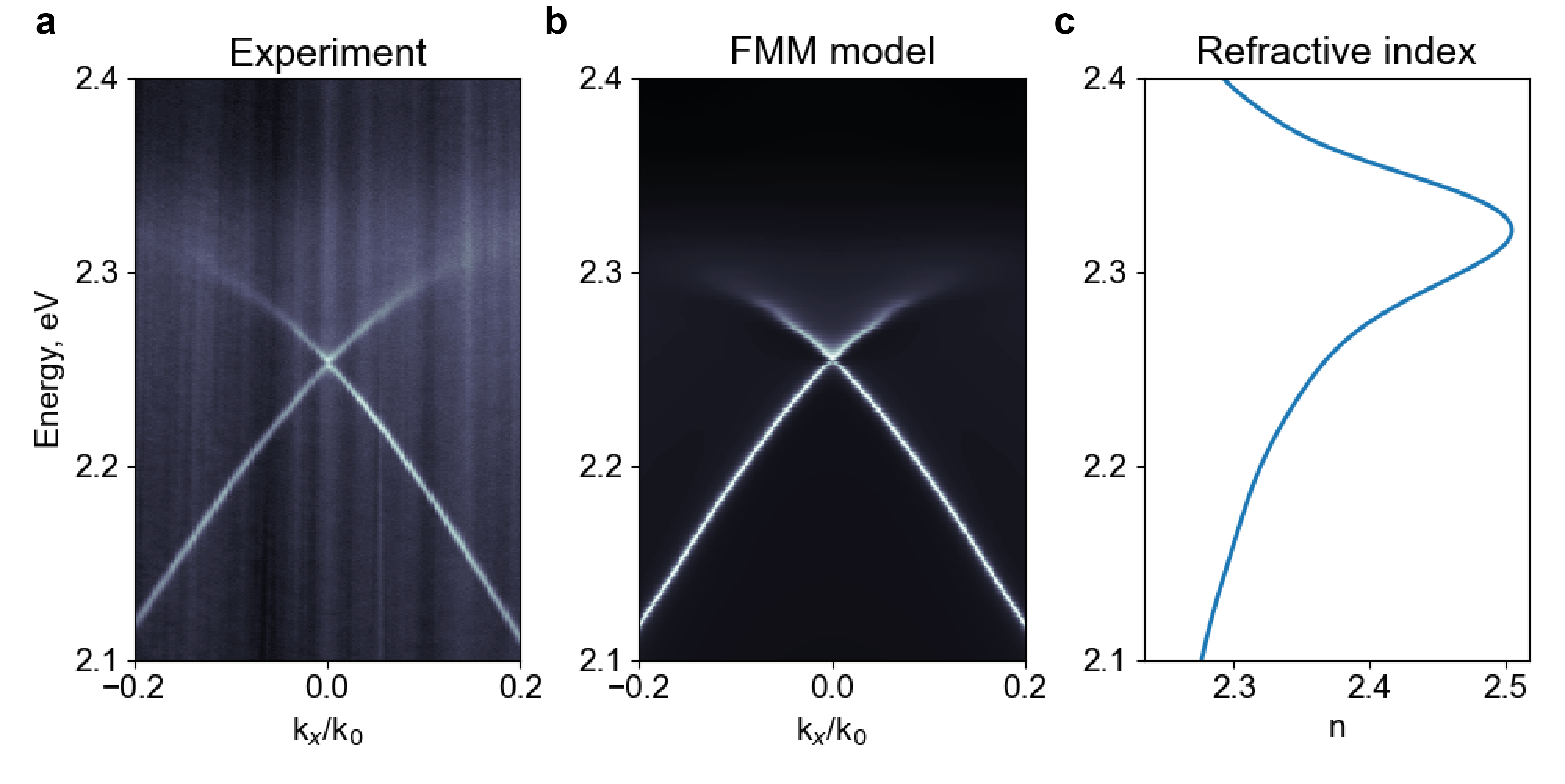}}
\renewcommand\thefigure{S3}
\caption{(a) The measured angle-resolved reflectivity spectrum of the lasing sample. (b) The calculated angle-resolved spectrum by FMM, based on the PCS geometry parameters extracted from the AFM and SEM measurements. The calculations correspond to the experimental results well. (c) Refractive index dispersion of MAPbBr$_3$ thin film, measured by the ellipsometry method.\cite{alias2016optical}}
\label{figSexpFMM}
\end{figure}

\begin{figure}[t!]
\centering
\center{\includegraphics[width=1\linewidth]{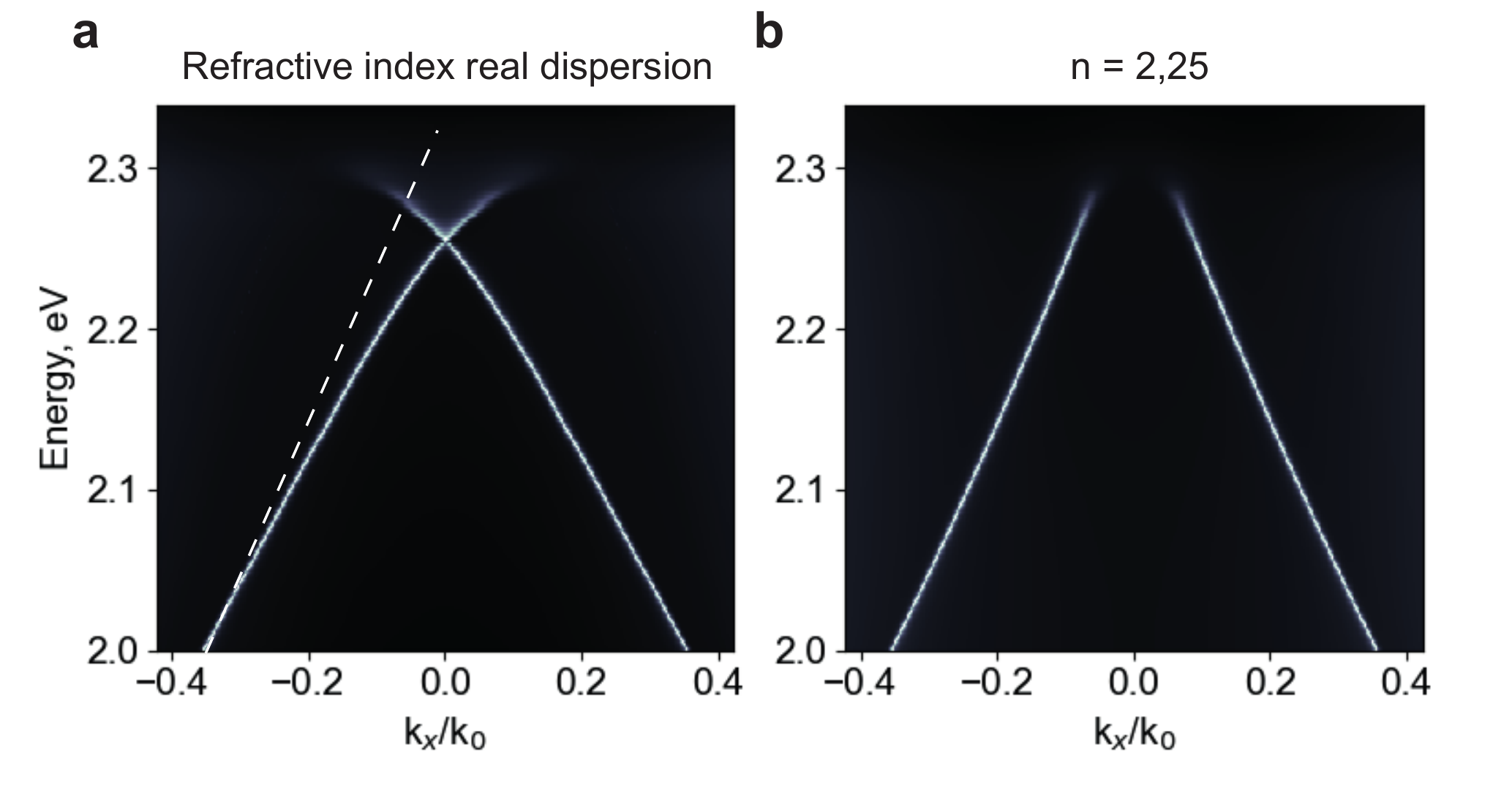}}
\renewcommand\thefigure{S4}
\caption{(a) Calculated FMM angle-resolved reflectance spectrum with real dispersion of the refractive index with the exciton resonance peak, shown in Fig.\ref{figSexpFMM}c. (b) Calculated FMM spectrum with constant refractive index, equal to 2,25. The value is chosen as the background for the exciton peak in the refractive index dispersion. With no excitonic peak leaky modes have approximately linear behavior, which is shown as a white dashed line in (a) }
\label{figS_FMMexciton}
\end{figure}

\begin{figure}[t!]
\centering
\center{\includegraphics[width=1\linewidth]{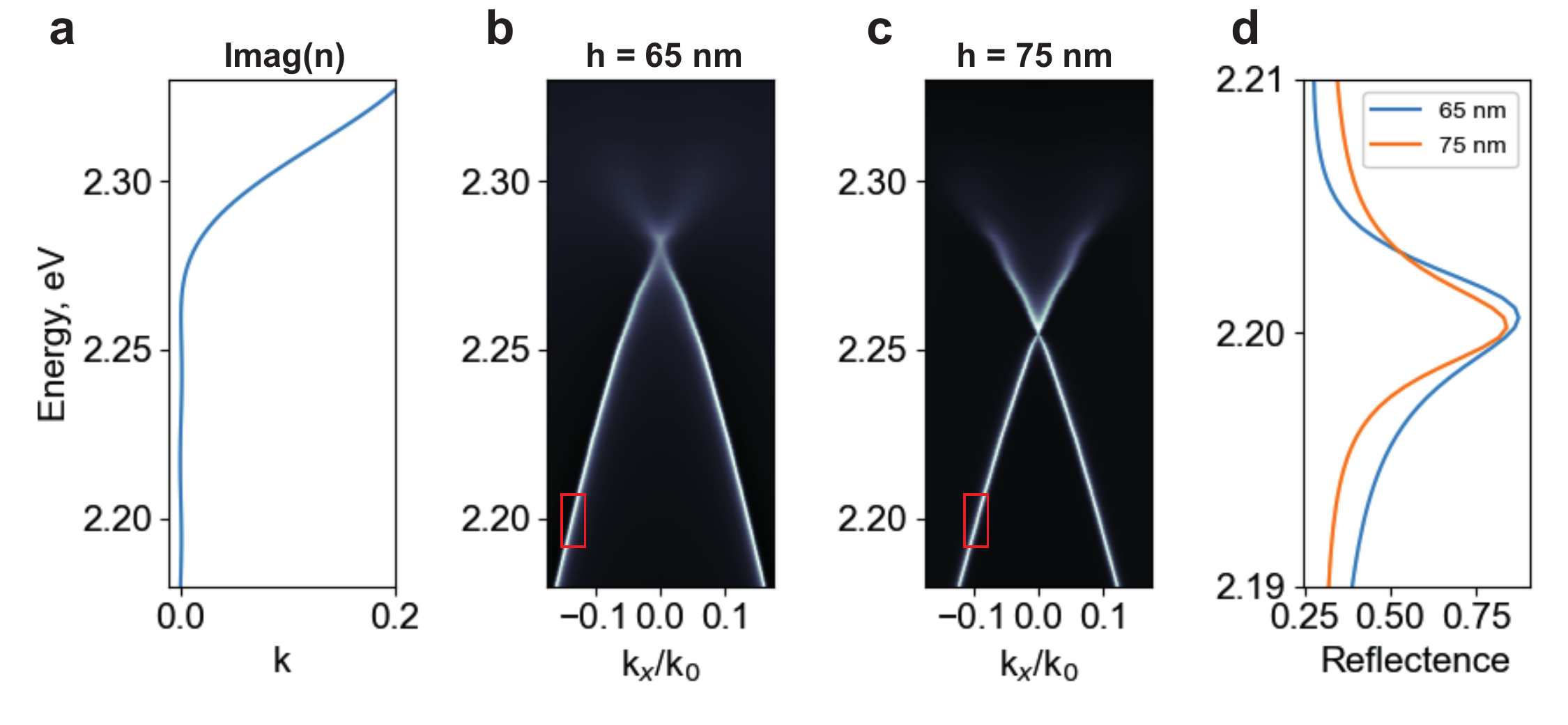}}
\renewcommand\thefigure{S5}
\caption{(a) Imaginary part of the MAPbBr$_3$ refractive index as a function of photon energy. (b,c) calculated reflectance spectra of the perovskite metasurface with a height of 65 and 75~nm, respectively. The spectral position of the resonance at the modes crossing point in the metasurface with $h = $~65~nm is located in the exciton absorption spectral region and therefore has enhanced non-radiative losses. (d) Extracted mode resonances around the energy of 2.2 eV, shown as red squares in panels b and c. The resonance of the 65~nm metasurface has a larger linewidth due to the higher relative modulation.}
\label{figS_FMM_diff_f}
\end{figure}

\begin{figure}[t!]
\centering
\center{\includegraphics[width=1\linewidth]{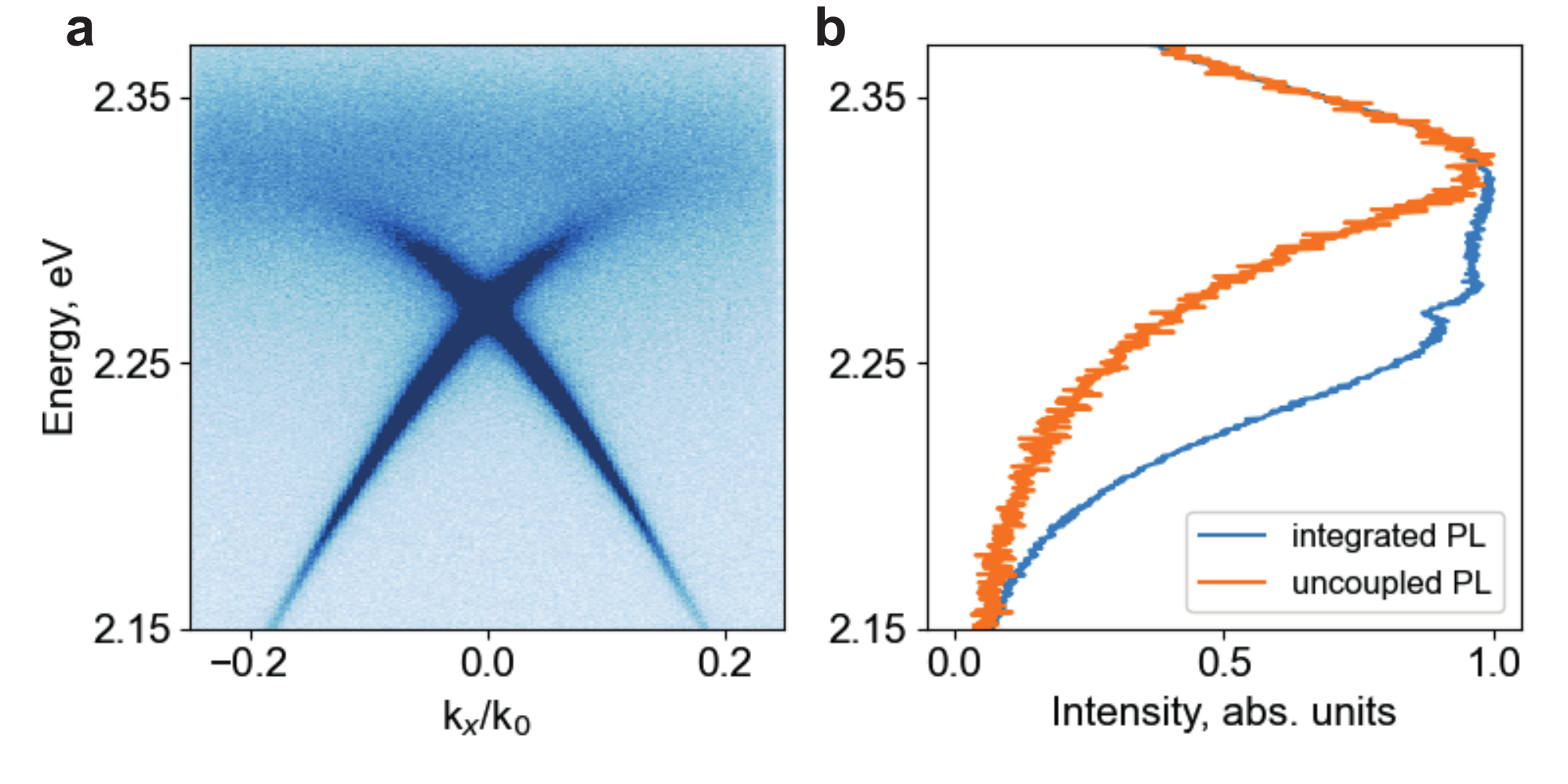}}
\renewcommand\thefigure{S8}
\caption{(a) Measured angle-resolved linear PL of the ASE sample with pseudocolor limits around uncoupled PL intensity. (b) Extracted integrated PL spectrum over $k_x/k_0$ from the data, shown in panel (a). And normalized uncoupled PL spectra, obtained by the integration over the $k_x/k_0$ in the range of values more than 0.2. It should be noted, that at the particular $k_x/k_0$ the intensity of the uncoupled PL is approximately 8 times lower than the intensity of the polariton mode PL and therefore is invisible in the spectra, shown in the main text}
\label{figSuncouplEXC}
\end{figure}

\begin{figure}[t!]
\centering
\center{\includegraphics[width=1\linewidth]{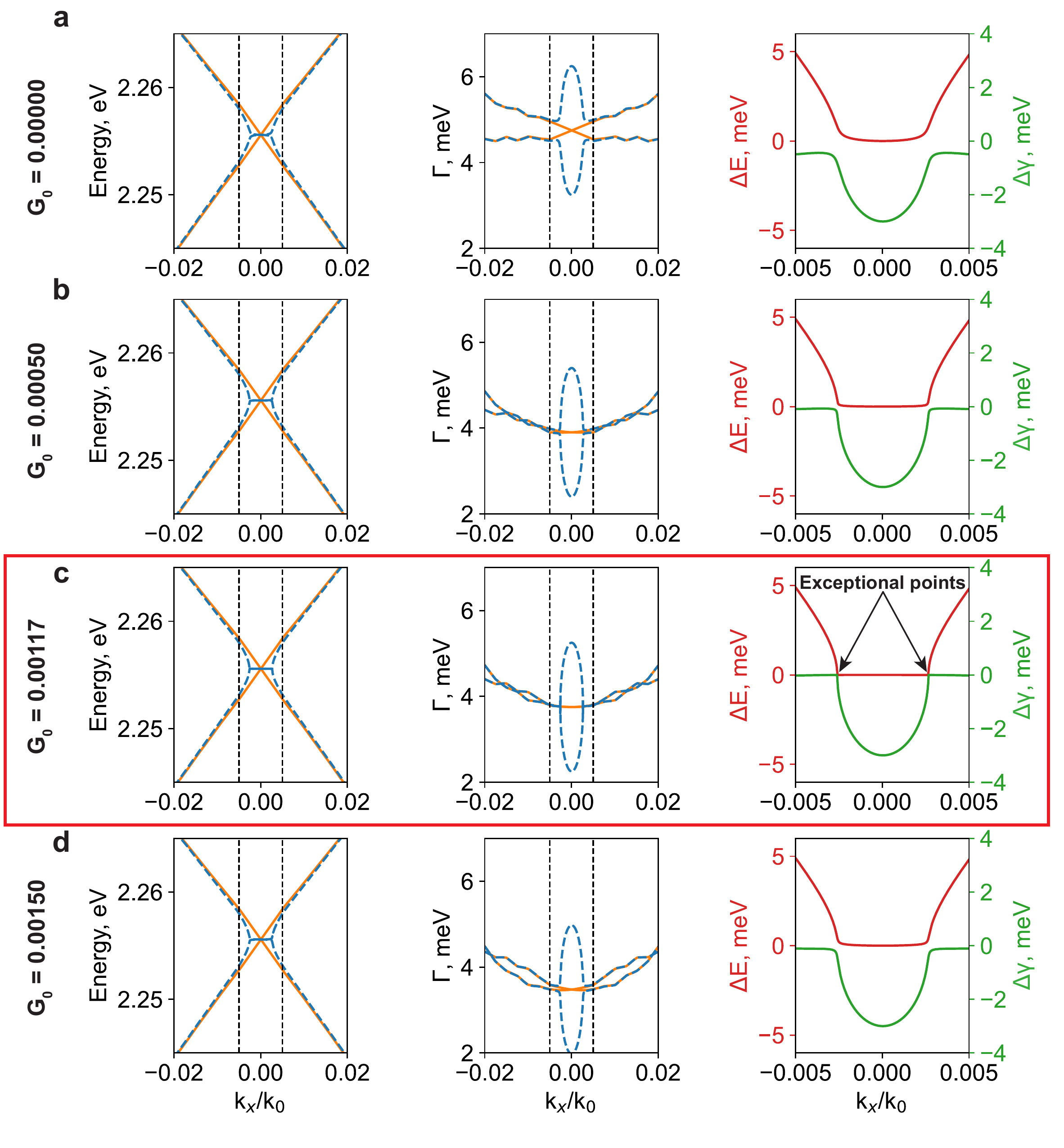}}
\renewcommand\thefigure{S10}
\caption{The result of the EPs modeling. Solid orange lines show the extracted real and imaginary parts of the extracted leaky modes. Dashed blue lines show the real (first column) and imaginary (second column) parts of the hamiltonian eigenvalues as a function of the wavenumber $k_x/k_0$. The red and green lines of the third column correspond to the real and imaginary parts of the difference between two eigenvalues respectively. (a-d) panels correspond to the different gain amplitude G$_0$, applied in the model, which is equal to 0, 0.0050, 0.00117, and 0.00150 respectively. The red rectangular shows the critical value to achieve the EPs.}
\label{figSanalysisEPsAppear}
\end{figure}

\begin{figure}[t!]
\centering
\center{\includegraphics[width=1\linewidth]{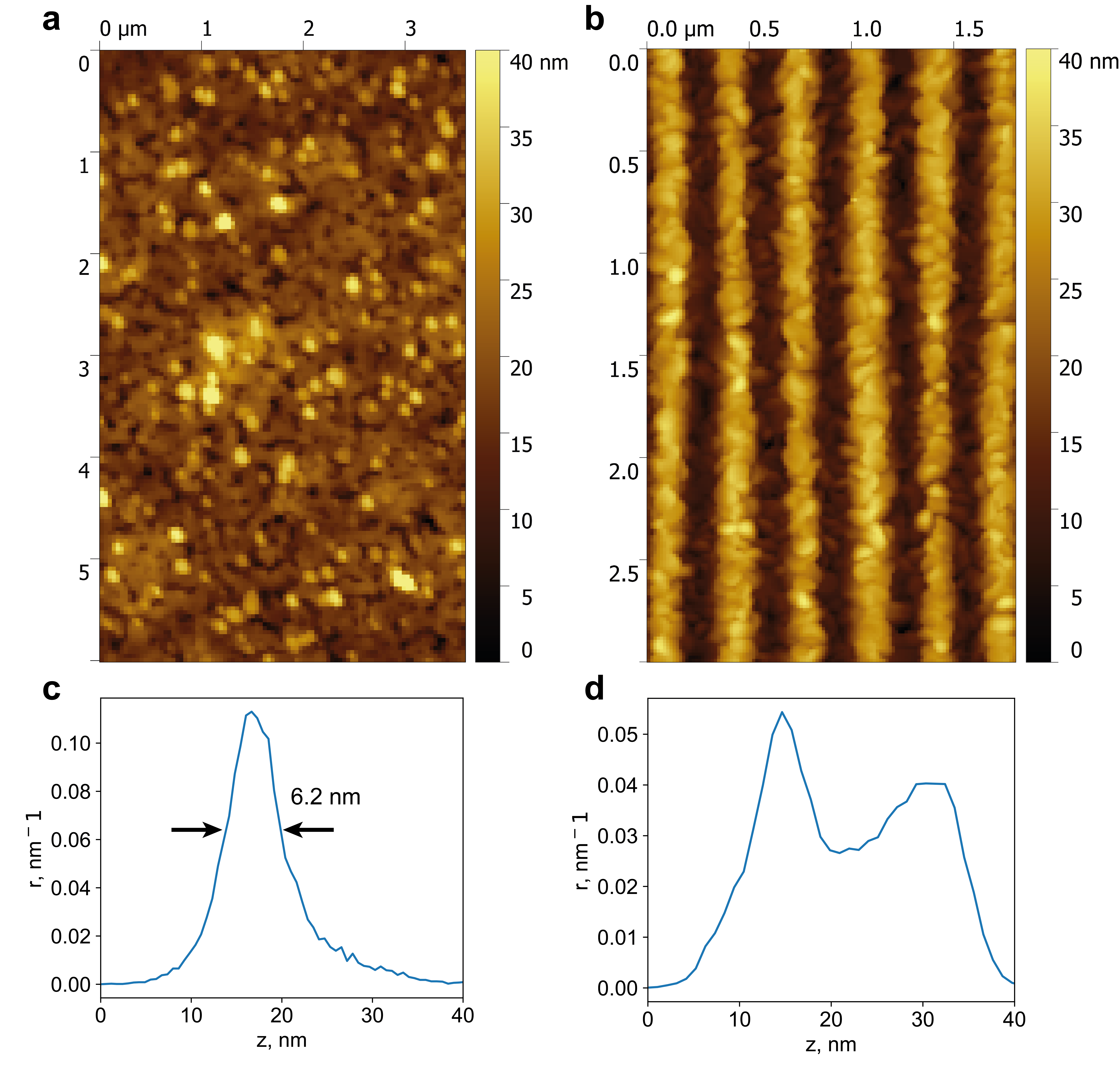}}
\renewcommand\thefigure{S11}
\caption{\red{ The measured morphology image with AFM of the pristine film (a) and metasurface (b). (c,d) Histograms of the pristine and imprinted samples, respectively. The RMS of the pristine film is estimated to be around 6.2~nm.} }
\label{figSanalysisEPsAppear}
\end{figure}

\begin{figure}[t!]
\centering
\renewcommand\thefigure{S12}
\center{\includegraphics[width=0.6\linewidth]{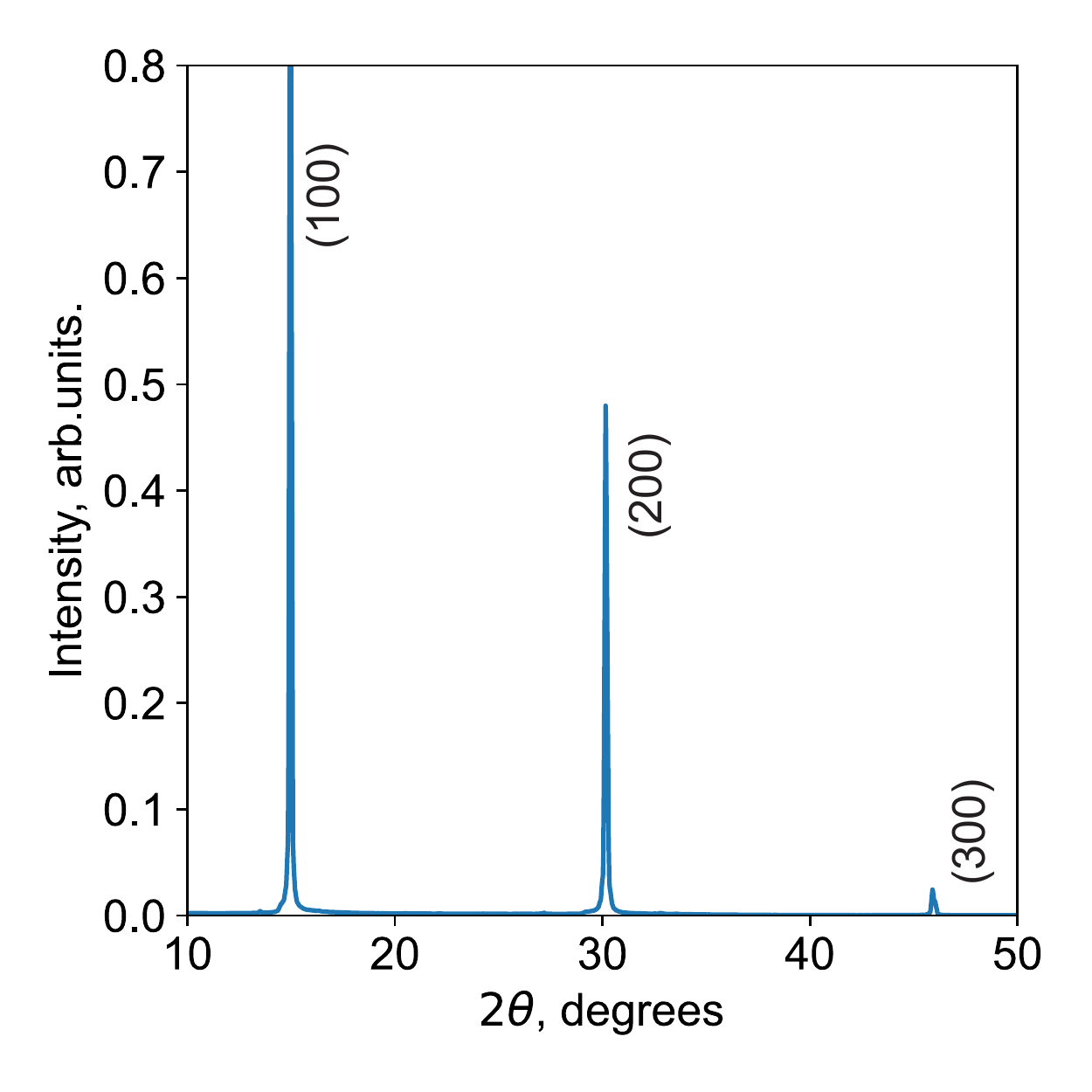}}
\caption{\red{The XRD pattern of the synthesized MAPbBr$_3$ thin film}}
\label{figSanalysisEPsAppear}
\end{figure}

\begin{figure}[t!]
\centering
\renewcommand\thefigure{S13}
\center{\includegraphics[width=0.7\linewidth]{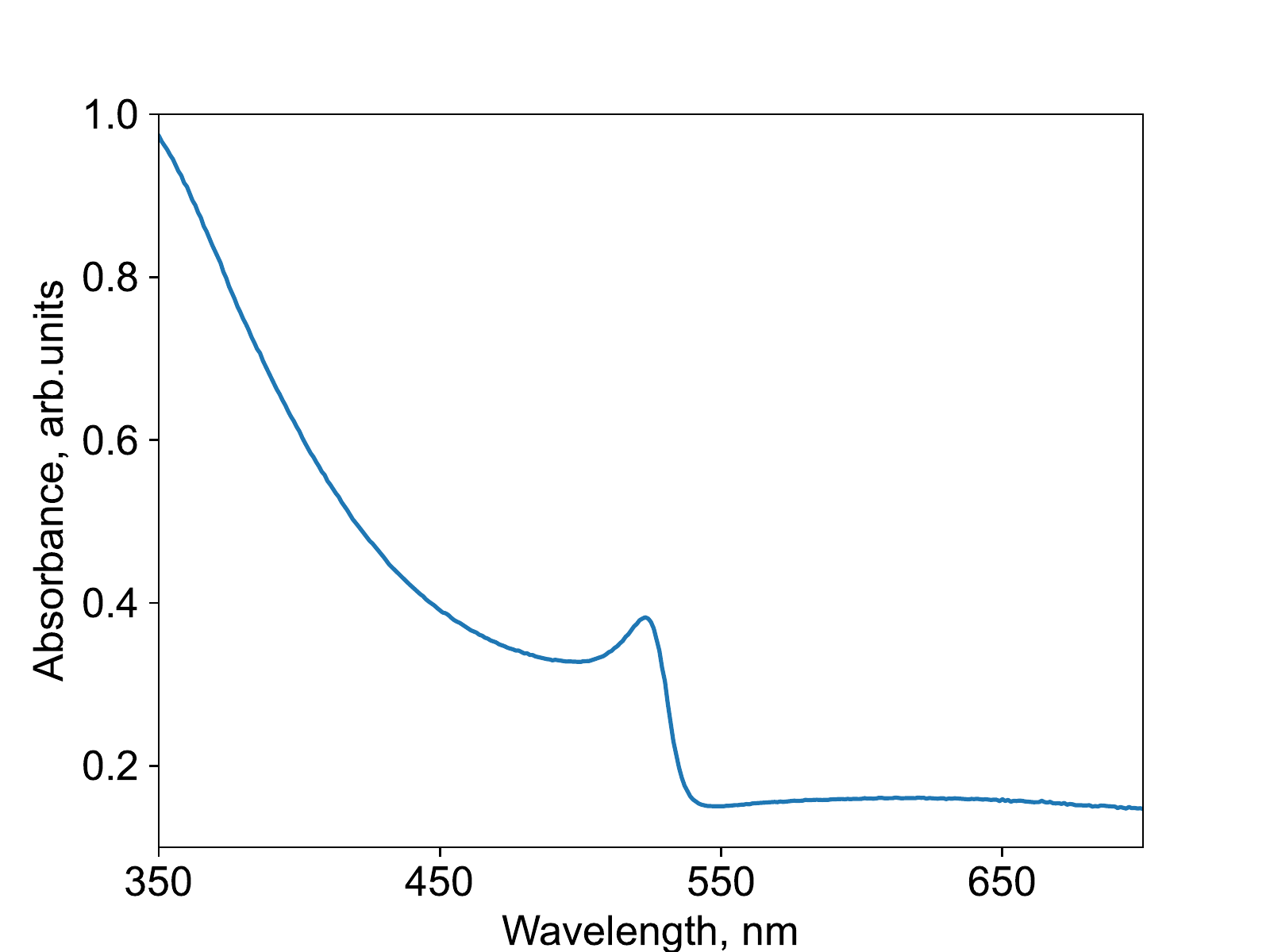}}
\caption{\red{The absorbance spectrum of the synthesized MAPbBr$_3$ thin film} }
\label{figSanalysisEPsAppear}
\end{figure}


